\documentclass[preprint2]{aastex}
\usepackage{amsmath}

\newcommand{\new}{}

\newcommand{\mj}{$M_{\mathrm{J}}$}
\newcommand{\rj}{$R_{\mathrm{J}}$}
\newcommand{\me}{$M_{\oplus}$}

\newcommand{\teq}{$T_{\rm eq}$}
\newcommand{\zpzs}{$Z_{\rm planet}/Z_{\rm star}$}
\newcommand{\zpl}{$Z_{\rm planet}$ }
\newcommand{\zs}{$Z_{\rm star}$}

\newcommand{\cp}{\citep}
\newcommand{\ct}{\citet}

\begin{document}
\title{The Mass-Metallicity Relation for Giant Planets}
\author{Daniel P. Thorngren\altaffilmark{1}, Jonathan J. Fortney\altaffilmark{2}, Ruth A. Murray-Clay\altaffilmark{2}, and Eric D. Lopez\altaffilmark{3}}
\altaffiltext{1}{Department of Physics, University of California, Santa Cruz, USA}
\altaffiltext{2}{Department of Astronomy and Astrophysics, University of California, Santa Cruz, USA}
\altaffiltext{3}{Institute for Astronomy, Royal Observatory Edinburgh, University of
Edinburgh, Blackford Hill, Edinburgh, UK}
\date{}
\begin{abstract}
Exoplanet discoveries of recent years have provided a great deal of new data for studying the bulk compositions of giant planets.  Here we identify 47 transiting giant planets ($20 M_\oplus < M < 20 M_{\mathrm{J}}$) whose stellar insolation is low enough ($F_* < 2\times10^8\; \text{erg}\; \text{s}^{-1}\; \text{cm}^{-2}$, or roughly $T_\text{eff} < 1000$) that they are not affected by the hot Jupiter radius inflation mechanism(s).  We compute a set of new thermal and structural evolution models and use these models in comparison with properties of the 47 transiting planets (mass, radius, age) to determine their heavy element masses.  A clear correlation emerges between the planetary heavy element mass $M_z$ and the total planet mass, approximately of the form $M_z \propto \sqrt{M}$.  This finding is consistent with the core accretion model of planet formation.  We also study how stellar metallicity [Fe/H] affects planetary metal-enrichment and find a weaker correlation than has been previously reported from studies with smaller sample sizes.  We confirm a strong relationship between the planetary metal-enrichment relative to the parent star \zpzs\ and the planetary mass, but see no relation in \zpzs with planet orbital properties or stellar mass.  The large heavy element masses of many planets ($>50$ \me) suggest significant amounts of heavy elements in H/He envelopes, rather than cores, such that metal-enriched giant planet atmospheres should be the rule.  We also discuss a model of core-accretion planet formation in a one-dimensional disk and show that it agrees well with our derived relation between mass and \zpzs.
\vspace{1 cm}\end{abstract}
\maketitle

\section{Introduction}\label{introduction}

Giant planets do not directly take on the composition of their parent stars.  If some flavor of core-accretion formation is correct \cp{Pollack1996}, then a seed core of $\sim 5-10$\me\ of ice/rock must build up first, which begins accreting nebular gas.  The gas need not share the exact composition of the parent star, due to condensation and migration of solids within the disk \cp{Lodders2009,Oeberg2011}.  The growing giant planet accretes gas but is also bombarded by planetesimals, which may add to the core mass or dissolve into the growing H/He envelope.  The amounts and variety of heavy elements (metals) accreted by the growing planet will depend on its formation location, formation time, disk environments sampled, and whether it forms near neighboring planets, among many other factors.  A planet's present-day composition is our indirect window into its formation process.

\new{The observed atmospheric composition of fully convective giant planets reflect the mixing ratios of heavy elements within their whole H/He envelope (with some caveats -- see \S \ref{distribution}).}  Spectroscopy of the Solar System's giants points to enhancements in the mixing ratio of carbon (as seen in CH$_4$) of $\sim$4, 10, 80, and 80, in Jupiter, Saturn, Uranus, and Neptune, respectively \cp{Wong2004,Fletcher2009}.  The \emph{Galileo Entry Probe} found that Jupiter's atmosphere is enhanced in volatiles like C, N, S, and noble gases by factors of 2-5 compared to the Sun \cp{Wong2004}. 

Remarkably, one can \new{make inferences about} the \emph{bulk} metallicity, or heavy element enhancement, \zpl /$Z_{\rm Sun}$, of a Solar System giant planet by measuring only its mass and radius.  \new{These results are consistent with and refined by measurements of their gravitational moments.}  By comparing to structure models, it is straightforward to infer from mass and radius alone that Jupiter and Saturn are both smaller and denser than they would be if they were composed of solar composition material \cp{Zapolsky1969,Fortney2007}. Therefore, we can infer they are enhanced in heavy elements.  As we describe below, this leads to an immediate connection with transiting giant planets, where mass and radius can be accurately measured.

\begin{figure}[t!]
    \centering
    \plotone{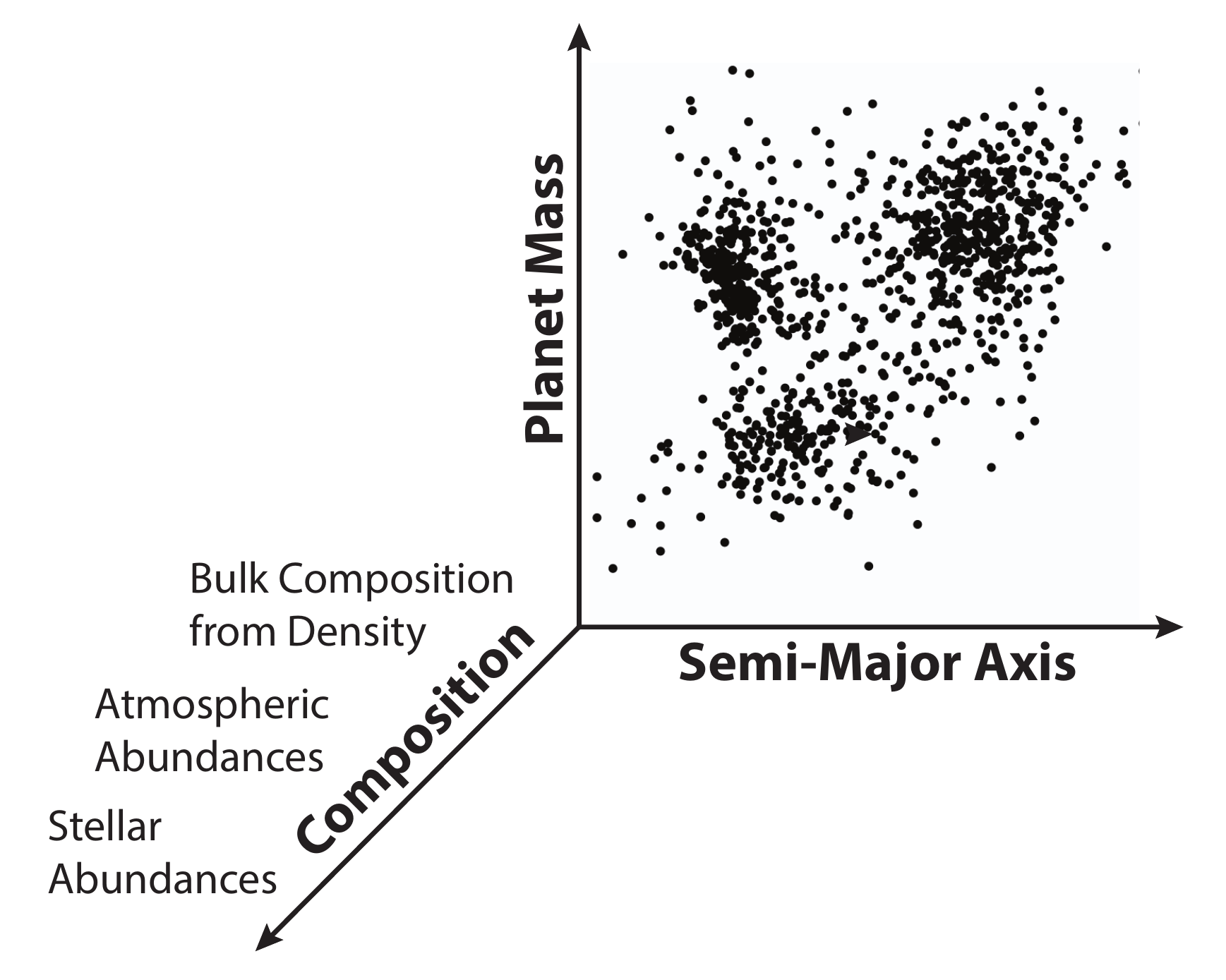}
    \caption{A highly schematic and simplified view of planes useful for understanding planet formation.  The main purpose of this study is to provide planetary composition information to inform planet formation models.}
    \label{plane}
\end{figure}

Knowledge of the heavy element enrichment of our Solar System's giant planets has led to dramatic advances in our understanding of planet formation.  Models of the core-accretion model of planet formation have advanced to the point where they can match the heavy element enrichment of each of the Solar System's giant planets \cp{Alibert2005,Lissauer2009,Helled2014}.  However, we are only beginning to attain similar data for exoplanets, which will provide a critical check for planet formation models over a tremendously larger phase space.  Such constraints are particularly important for comparison with planetary population synthesis models that aim to understand the processes of core formation, H/He envelope accretion, and planetary migration, in diverse protoplanetary environments \cp{Ida2004,Mordasini2014}.  Constraints on planet formation from exoplanetary systems have been almost entirely driven by data on the frequency of planets in the mass/semi-major axis plane.  A promising avenue is to move these comparisons into a complementary plane, that of planetary composition, either in bulk composition \cp{Guillot2006,Burrows2007,Miller2011,Mordasini2014}, or atmospheric composition \cp{Madhusudhan2011,Fortney2013,Kreidberg2014,Barman2015,Konopacky2013} as shown in a schematic way in Figure \ref{plane}.

The dramatic rise in the number of observed transiting exoplanets provides a unique opportunity.  With radii derived from transit observations and masses derived from radial-velocity or transit-timing variation measurements, we get especially detailed information about these objects.  This gives us a measured density, and therefore some rough information about their bulk composition.  A more advanced analysis uses models of planet structural evolution (e.g. \cite{Fortney2007}) to constrain the quantity of heavy elements.

Most of the giant planets we have observed are strongly irradiated hot Jupiters, whose radii are inflated beyond what models predict.  Much effort has been put into understanding this discrepancy.  A thorough discussion is outside the scope of this article, but the various proposed inflation mechanisms are extensively reviewed in \cite{Fortney2010}, \cite{Baraffe2014}, and \cite{Weiss2013}.  Unfortunately, without a definite understanding of the inflation process this acts as a free parameter in modeling: the inflationary effect enlarges a planet and added heavy elements shrink it, resulting in a degeneracy that inhibits our ability to obtain useful composition constraints.  Still, work has been done to use models to address a the star-planet composition connection, using plausible assumptions about the effect, as a heat source \citep{Guillot2006} or as a slowed-cooling effect \citep{Burrows2007}.  Both studies saw an increase in planet heavy element mass with stellar metallicity.

\begin{figure}[b!]
    \centering
    \plotone{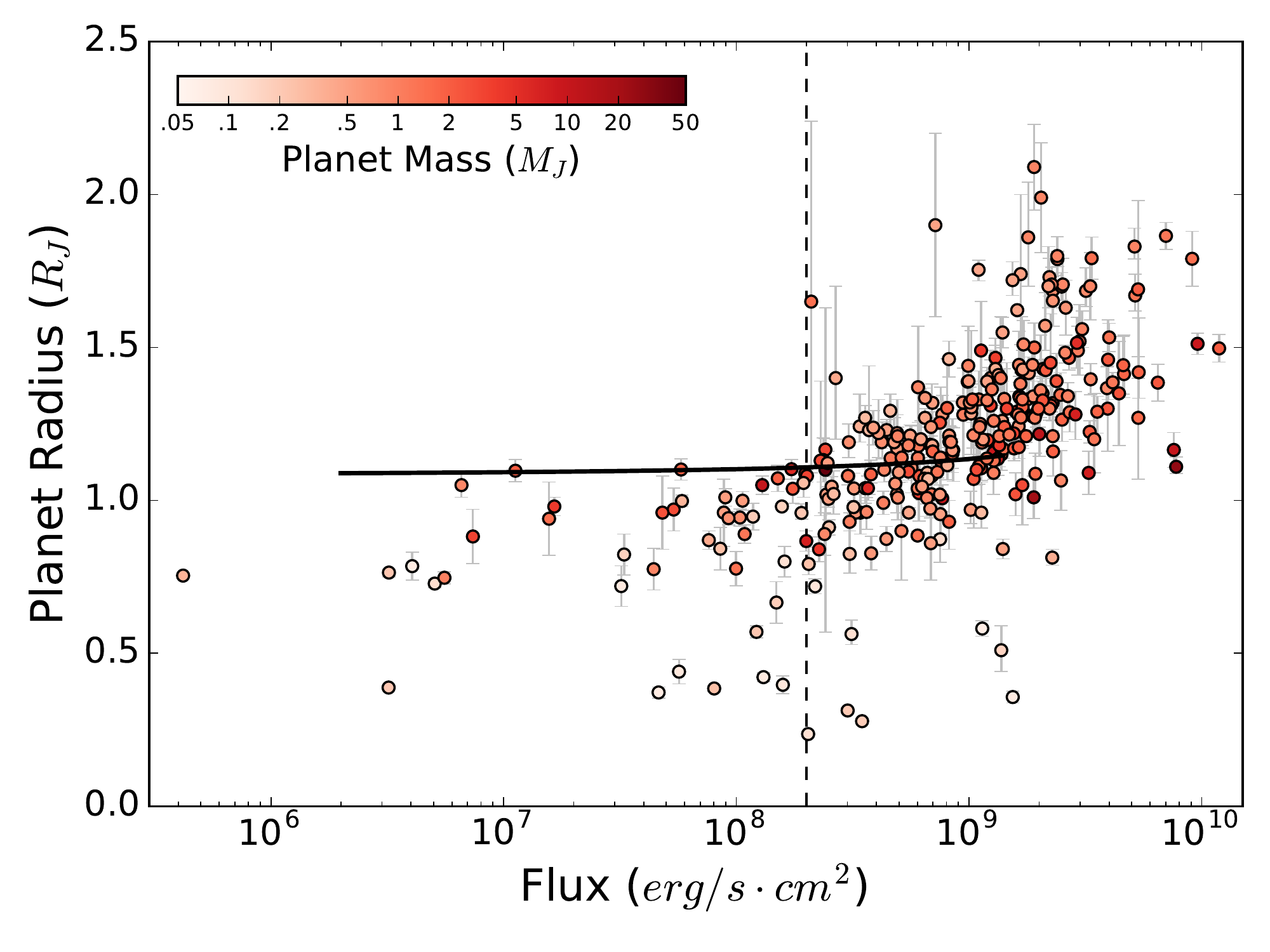}
    \caption{Planetary radii for observed planets against stellar insolation.  The black line is an 4.5 Gyr., 1 \mj\; pure H/He object, roughly the maximum radius for older, uninflated planets.  The dashed vertical line marks the flux cutoff we use to identify the cool, uninflated giants.}
    \label{flux}
\end{figure}

A promising avenue of investigation are the sample of transiting exoplanets which are relatively cool.  Planets that receive an incident flux below around $2\times10^8\; \text{erg}\; \text{s}^{-1}\; \text{cm}^{-2}$ (\teq $\lesssim$ 1000 K) appear to be non-inflated \citep{Miller2011,Demory2011}, obviating the need for assumptions about that effect.  Figure \ref{flux} shows this threshold.  \cite{Miller2011} (hereafter referred to as MF2011) studied these planets, finding correlations in the heavy element mass with planetary mass and stellar metallicity.  In particular, they noted a strong connection between the relative enrichment of planets relative to their parent stars (\zpzs) and the planet mass.  However, that study was limited by a small sample size of 14 planets.

In our work that follows, we consider the set of cool transiting giant planets, now numbering 47, and compare them to a new grid of evolution models to estimate their heavy element masses, and we include a more sophisticated treatment of the uncertainties on our derived planetary metal-enrichments.  We then examine the connections between their mass, metal content, and parent star metallicity.

\section{Planet Data and Selection}\label{planetData}
Our data was downloaded from the Extrasolar Planets Encyclopedia \citep[exoplanets.eu,][]{Schneider2011}) and the NASA Exoplanet Archive \citep{Akeson2013}.  These data were combined, filtered (see below) and checked against \new{sources} for accuracy.  Some corrections were needed, mostly in resolving differing values between \new{sources}; we aimed to consistently use the most complete and up-to-date \new{sources}.  We tried to include all known planets who met the selection (even if some of their data was found in the literature and not the websites).  Critically, we use data from the \new{original sources}, rather than the websites (see Table 1).

For our sample, we selected the cool giant planets that had well-determined properties.  Typically, this means they were the subject of both transit and radial velocity studies.  The mass and radius uncertainties were of particular importance.  Planets needed to have masses between $20 M_\oplus$ and $20 M_J$, and relative uncertainties thereof below 50\%.  Our sample's relative mass uncertainties were typically well below that cutoff, distributed as $10_{-5.7}^{+12.8}\%$.  We also constrained relative radius uncertainty to less than 50\%, but again saw values much lower ($5.0^{+4.6}_{-2.5}\%$).  Both uncertainty cuts were made to eliminate planets with only rough estimates or upper limits for either value.

As discussed in \S\ref{introduction}, we used a $2\times10^8\; \text{erg}\; \text{s}^{-1}\; \text{cm}^{-2}$ upper flux cutoff to filter out potentially inflated planets.  Consequently, candidates needed to have enough information to compute the \new{time-averaged} flux: stellar radius and effective temperature (for both stars, if binary), semi-major axis, and eccentricity.  In addition, we needed measured values for the stellar metallicities in the form of the iron abundance [Fe/H].  These tended to have fairly high uncertainties, and were a major source of error in our determination of \zpzs .

Determining the age of a planet is necessary to use evolution models to constrain its metal content; unfortunately, this is often a difficult value to obtain.  Our ages come from stellar ages listed in the literature, typically derived from a mixture of gyrochronology and stellar evolution models.  These methods can produce values with sizable uncertainties, typically given as plausible value ranges.  We treat these as flat probability distributions (a conservative choice), and convert values given as Gaussian distributions to 95\% confidence intervals.

Because planets do most of their cooling, and therefore contraction, early in their lives \citep[see][for a discussion]{Fortney2007}, large uncertainties in age are not a major obstacle in modeling older planets.  For planets who may be younger than a few Gyr, we cannot rule out that they are very metal-rich, but young and puffy, as the two effects would cancel out.  We account for this in our analysis; planets which may be young consequently exhibit higher upper bound uncertainties in heavy element mass.  For Kepler-16 (AB)-b, Kepler-413 (AB)-b, and WASP-80 b, no age was given, so we used a range of .5-10 Gyr.  This is reasonable because it represents the possibilities of the planets being either young or old, and because the age was a second-order effect on our heavy element assessment (after mass and radius)\new{as seen in Figure \ref{evolution}}.

\section{Models}
\begin{figure}[t!]
    \centering
    \plotone{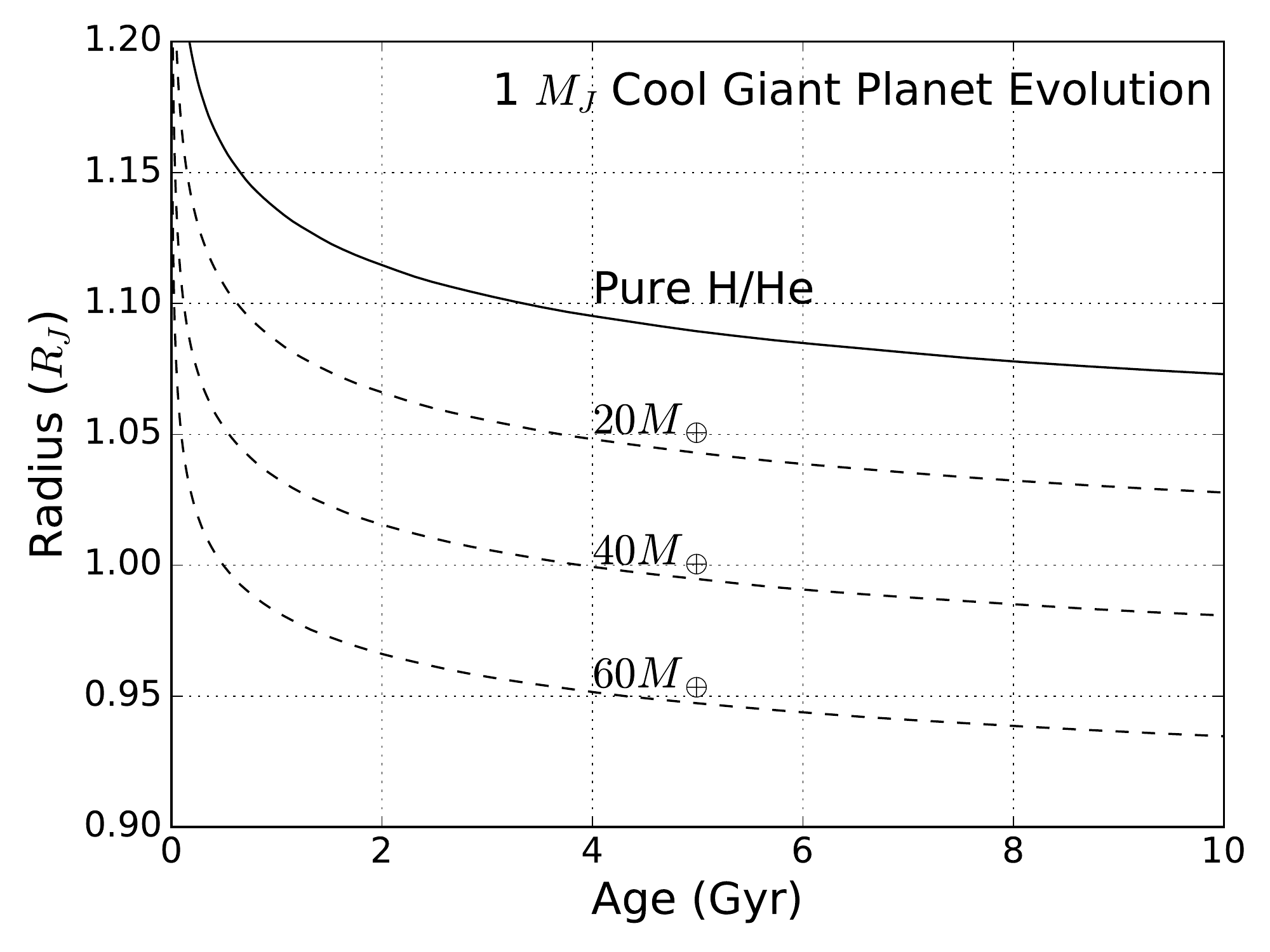}
    \caption{The radius evolution over time of 1 $M_J$ planets with low stellar insolation ($1\times10^7\; \text{erg}\; \text{s}^{-1}\; \text{cm}^{-2}$) containing various quantities of metal.  \new{The plot demonstrates that our models have reasonable behavior, shrinking with age and heavy-element enrichment.  The amount of metal present has an effect on the radius substantial enough to be observed in exoplanet populations.}}
    \label{evolution}
\end{figure}

We created one-dimensional planetary models consisting of an inert core composed of a 50/50 rock-ice mixture, a homogenous convective envelope made of a H/He-rock-ice mixture, and a radiative atmosphere as the upper boundary condition.  The models are made to satisfy the equations of hydrostatic equilibrium, mass conservation, and the conservation of energy.
\begin{equation}
\frac{\partial P}{\partial m} = -\frac{G m}{4 \pi r^4}
\end{equation}

\begin{equation}
\frac{\partial r}{\partial m} = \frac{1}{4 \pi r^2 \rho}
\end{equation}

\begin{equation}
\label{energyConservation}
\frac{\partial L}{\partial m} = - T \frac{\partial S}{\partial T}
\end{equation}
Equation \ref{energyConservation} is not used in the core, where luminosity is neglected.  Structures are initially guessed, then iteratively improved until these conditions are met to within sufficiently small error.  This computation was done in a new Python code created for this study.  Its advantages are its relative speed (a 6 million planet grid takes $<1$ hour to create) and its ability to easily try different compositional structures (e.g. heavy elements in the core vs.~envelope) and/or equations of state.  \new{As a simple visual diagnostic, Figure \ref{evolution} shows the output of our models.}

For our atmosphere models, we interpolate on the solar metallicity grids from \cite{Fortney2007}, as was done in several other \new{works} (including \cite{Miller2009} and \cite{Lopez2014}.  With these models we compute the intrinsic luminosity ($L$) of a planet model (see equation \ref{energyConservation}), which describes the rate that energy escapes from the interior, at a given surface gravity and isentropic interior profile.  These grids are then used to determine the rate of entropy change in the envelope, and the contraction history of a given model planet.  The initial entropy is not important; reasonable initial values typically all converge within a few hundred million years \citep{Marley2007}.  As such we do not need to consider if planets form in a hot or cold start scenario.  Following \ct{Miller2009}, we include the small extension in radius due to the finite thickness of the radiative atmosphere.

A fully self-consistent treatment of the atmosphere would use a range of metal-enriched atmosphere grids to be interpolated to yield consistency between atmospheric metallicity and H/He envelope metal mass fraction.  Metal-enriched grids would tend to slow cooling and contraction \cp{Burrows2007} and yield \emph{larger} heavy element masses than we present here.  However, given uncertainties in the upper boundary condition in strongly irradiated giant planets \new{(see, e.g., \cite{Guillot2010} for an analytic analysis,\cite{Spiegel2013} for a review of the chemical and physical processes at play, and \cite{Guillot2002} for an application of a 2D boundary condition)}, and our uncertainty in where the heavy elements are within the planets (core vs.~envelope) it is not clear if such an expanded study is yet warranted.

In creating our models, we had several important factors to consider: where the heavy elements are within the planet, what the composition of these heavy elements is, and how to treat the thermal properties of the core.  These questions do not have any clearly superior or well-established solutions.  The uncertainties from them, however, were overshadowed by the large uncertainties from available values for mass, radius and (to a lesser extent) age.  Therefore, in designing our models, we chose plausible solutions instead of attempting to incorporate all modeling uncertainties into our results.

\subsection{Heavy Element Distribution}\label{distribution}
Hydrogen-Helium mixtures are more compressible than typical heavy elements, and so models with heavy elements in a core tend to be larger than models with the same heavy element mass mixed throughout the envelope \citep[see][for a discussion]{Baraffe2008}.  Conversely, modeling planets with pure heavy element cores and pure H/He envelopes requires the most heavy elements of any structure for a given total mass and radius.  For some planets in our sample we find this kind of model requires implausibly high metal-enrichment to explain a given radius (cores of several \emph{Jupiter masses} for extreme cases), and it is difficult to imagine how such massive cores could form.  As such, in our models here we distribute the heavy element mass by putting up to 10 $M_\oplus$ into a pure heavy element core, and then use linear mixing to put any remaining metal mass in the otherwise H/He envelope.  This allows us to consistently model both core-dominated low-mass planets and likely better-mixed massive planets.

\new{For our work, we are assuming a homogeneous, isentropic envelope.  In our solar system, however, at least some inhomogeneities must exist \citep{Chabrier1992}, such as helium phase separation in Saturn.  Layered semi-convective models are also consistent with structure models \citep[e.g.][]{Leconte2012}, though more work is needed to understand the origin and maintenance of such layers.  At a given metal mass, such structures suppress planetary heat loss, resulting in larger radii at a given age.  Therefore, our model would underestimate the heavy-element masses of such planets.  Such a model implies that cooler giants could be ``anomalously" inflated if they have the right compositional structure \cite{Chabrier2007}, but no such planets have been observed.  We conclude that our homogeneous model is an acceptable approximation, but look forward to future work in understanding how composition interacts with thermal evolution.}

\subsection{Equations of State} \label{EOS}
For hydrogen and helium we used the \cite{Saumon1995} equation of state (EOS), with a solar ratio of hydrogen to helium ($Y=0.27$).  For envelopes with metals mixed in, we used additive volumes to adjust the equation of state.  Our choice for the metal EOS was also important -- denser materials like iron produce noticeably smaller planets (and therefore require a smaller \zpl to explain a given planet).  Olivine, a mineral whose EOS is commonly used to represent rock, is less dense than iron.  Water, used as a proxy for ices generally, is less dense still.  MF2011 showed that changing the metal composition produces differences on the order of 20\%, consistent with our models.  We chose to use a 50-50 rock-ice mixture, using the \cite{Thompson1990} ANEOS equation of state.  This would overestimate \zpl if the metal were actually iron-dominated, but this seems unlikely to occur commonly in giant planets, which are typically expected to form near the snow line \citep{Ida2004}.

\section{Analysis} \label{Analysis}

To apply our models to an observed planet, \new{we take draws from the probability distributions implied by the measurement uncertainties in mass, radius, age.  Each draw therefore consists of a mass, radius and age for one planet.}  \new{The probability distributions are normal}, except for the age, which is conservatively modeled as a flat range (see \S \ref{planetData}).  \new{For each draw we compute the inferred heavy-element mass.  By making many (10,000) draws we have an estimate of the range of heavy-element masses consistent with observations.  This procedure is done for each planet in the sample.}

\begin{figure}
    \centering
    \plotone{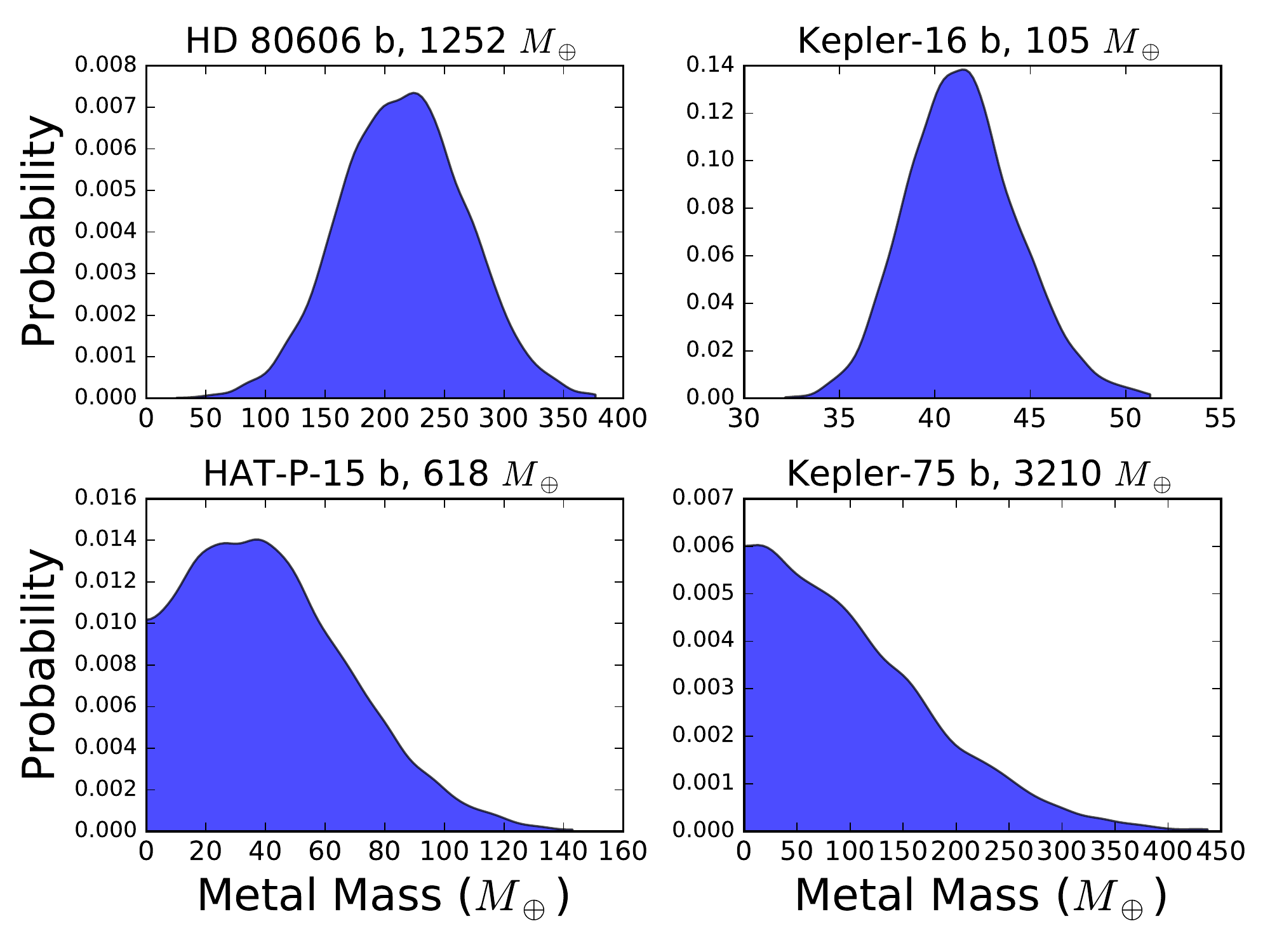}
    \caption{Plots of the inferred heavy element masses for four giant planets, using a Gaussian KDE of 10,000 samples each.  The top two, HD 80606b and Kepler-16b, have distributions typical of the sample.  HAT-P-15b is one of the six planets for which more than a $1 \sigma$ portion of the distribution extends below the pure H/He limit.  Each of these planets has had their lower error bars extended to zero, changes marked with a "$\dagger$" on Table 1.  Kepler-75b is the single planet for which only an upper limit could be determined; its RMS heavy element mass (134 $M_\oplus$) is reported as the upper error.}
    \label{rephist}
\end{figure}

The \new{resulting distributions} were single peaked and roughly Gaussian overall (see Figure \ref{rephist}). \new{For some of our planets, the uncertainty was dominated by the mass or radius, as evidenced by a correlation among the draws for each individual planet.  Mass error more commonly dominated the uncertainty at high $Z_\textrm{planet}$, and radius error at low $Z_\textrm{planet}$.}  Our reported values show the \new{marginal} mean of each distribution, with upper and lower uncertainties computed as the \new{RMS deviation from the mean from draws above the mean and below the mean, respectively}.  These represent the data reasonably well, but care should be taken not to overlook the correlation with input variables.  For our part, we do all computation directly on the samples (see below), and report the uncertainties in the resulting distributions.  

Some planets whose \zpl values were clustered near zero (pure H/He) or one (pure ice/rock) generated \new{draws} which could not be recreated in our models.  \new{This occurred if (for example) the draw was randomly assigned a low value for mass and a high value for its radius, such that it was larger than an analogous pure H/He object.}  These \new{draws} were discarded, but we noted how often this occurred.  For the six planets where this occurred 15.8\% - 50\% of the time (where at least a 1 $\sigma$ tail is outside the valid range), the error-bar on that side was adjusted to a \zpl of exactly 0 or 1, as appropriate.  These are marked with a $\dagger$ in Table 1.  For the massive H/He dominated planet Kepler-75b, this occurred more than half the time, so its heavy element derived values are listed as upper limits.

To conduct the regressions described in our results (\S \ref{Results}), we use \new{a linear Bayesian regression on the log of the values of interest, using uninformative priors on the slope and intercept.  The likelihood for the regression was
\begin{equation}
    P(\vec{y}|\bf{X},\vec{\beta},\sigma^2) = \text{Normal}(\bf{X}\vec{\beta},\sigma^2 \bf{I})
\end{equation}
where $\vec{y}$ is the vector of y points, $X=[\vec{1},\vec{x}]$ is the matrix of covariates, $\vec{\beta}$ is the coefficients vector (essentially [b,m] from $y=mx+b$), and I is the identity matrix.  Using the standard noninformative prior $P(\beta,\sigma^2|\bf{X}) \propto 1/\sigma^2$ and the distribution of $\vec{y}$ and the covariate $\vec{x}$ as  $P(\vec{y},\vec{x})=\prod_i P_i(y_i,x_i)$ we derived the full conditional distributions and implemented a Gibbs sampler.  For each fit we initialized with the classical fit, and had a burn in of 1000 steps and thinned the results by keeping only every tenth step to ensure the results were well-mixed.  Our fits were performed in logspace so that they were effectively power-law fits.}

We considered the possibility that mass and radii observations are not independent.  If they are correlated, then sampling more directly from observational data (or posteriors thereof) could improve the uncertainties in our heavy element masses.  Such an operation would need to be careful not to \new{extract more information than the data actually provide}.  \cite{Southworth2007} describes a method of computing surface gravity which is more precise than directly using a planet's derived mass and radius.  They instead use orbital parameters, the planet's radius, and the stellar reflex velocity $K$.  These are closer to the observed quantities, and so avoid unnecessarily compounding uncertainties.  We used Southworth's formula as a proxy to estimate how much improvement in uncertainty we might see from such an approach.  We found that most of our planets exhibited only modestly better gravity estimates (though a handful were substantially improved, such as HAT-P-18 b).  As such, we determined that trying to work more directly from observation data would not be worthwhile for our set of planets.  Still, we suggest that studies examining individual planets should consider this approach.

\subsection{Modeling Uncertainty}
\new{In the preceding sections, we listed a number of possible sources of modeling uncertainty.  We will now argue that these uncertainties, while present, do not significantly affect our results, especially the fits described in \S \ref{Results}.  To the extent that they are affected, the error should be concentrated in the coefficient of our fits (not the power) because to first order these uncertainties would affect all planets equally.}

\begin{figure}[t!]
    \centering
    \plotone{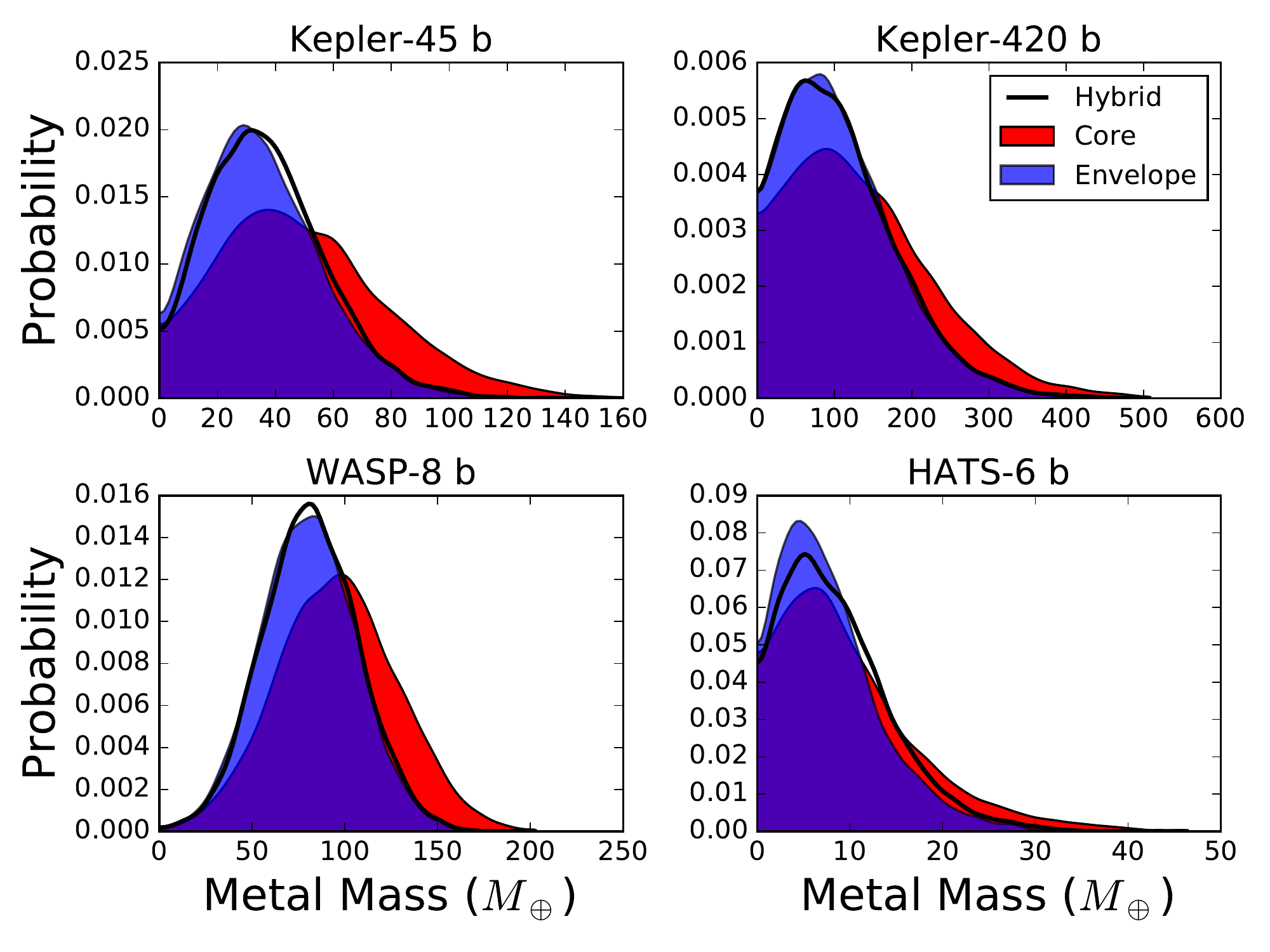}
    \caption{\new{A comparison of three Z distribution models of inferred heavy-element abundance for four of our sample planets using a Gaussian KDE.  Blue represents a fully homogenous envelope with no core, red is a model with all heavy-elements located in a central core, and the black line is our intermediate model.  The spread comes from observational uncertainties on the mass, radius, and age of the planet (see \S \ref{Analysis}).  Although the model can have a significant effect on the inferred metal abundance, this  effect is much smaller than observational uncertainties.}}
    \label{zModelComparison}
\end{figure}

\new{First, we consider the effect of the heavy-element distribution on the structure and evolution of the planets.  Considering the two extreme cases, where the metal is either entirely confined to a core or entirely dissolved into a homogeneous envelope, we compared the resulting models of four representative giant planets against our preferred models.  As can be seen in Figure \ref{zModelComparison}, these different choices have a clear effect on the inferred metal abundance, but the effect is small compared to observational uncertainties.}

\new{To evaluate the effect of EOS uncertainty, we considered the difference between the Saumon-Chabrier H/He EOS that we used and the \cite{Militzer2013} EOS.  This EOS is computed from DFT-MD simulations which may be more accurate than the semi-analytic SCvH EOS.  We were unable to use it for this work because it only covers densities up to those found in roughly Jupiter-mass planets.  Figure 12 from \cite{Militzer2013} shows that for envelope entropies typical of older planets, the deviation in the resulting radius is about 10\%.  To match this, we might require as much as 15\% less metal. In practice the amount is somewhat lower due to next-order effects: e.g., smaller planets evolve slower (less surface area to emit from) and the metal EOS of the planets is unaffected by a H/He EOS change.}

\begin{figure}[t]
    \centering
    \plotone{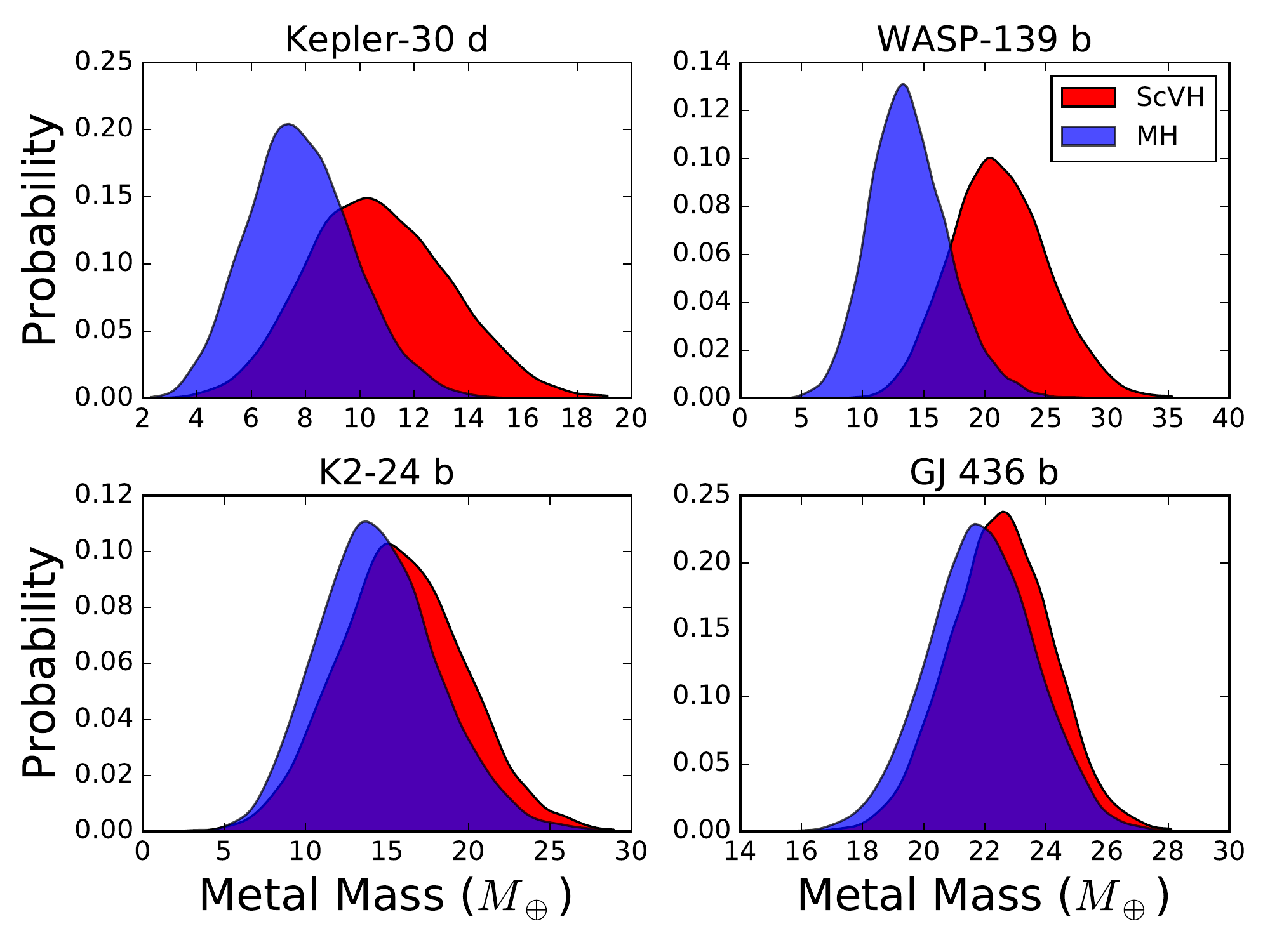}
    \caption{\new{A comparison of inferred heavy-element masses for four sample planets using the ScVH and \cite{Militzer2013} EOS' with a Gaussian KDE, assuming a core-only metal distribution.  Most planets are only modestly impacted, but our handful of very young planets like WASP-139 b and Kepler-30 d ($<1$ Gyr and $<3.8$ Gyr) are affected more strongly.}}
    \label{eosComp}
\end{figure}

\new{To quantify this, we derived inferred metal-masses of our planets using the Militzer-Hubbard EOS where possible.  The results for four planets are shown in Figure \ref{eosComp}.  This sample of planets is \emph{not} representative in mass because the Militzer-Hubbard EOS does not extend to high enough pressures to model super-Jupiters.  Most of the results were fairly similar (bottom row), but a few, generally young planets, exhibited more significant differences.  The choice of EOS matters, but is usually a next-order effect after observational uncertainties.}

\new{As a reference, we applied our model to Jupiter and Saturn.  Since these planets have well-determined properties that include some gravitational moments, we can use them as a test of our model's validity.  Our inferred heavy-element mass should resemble estimates from better-constrained models that make use of these gravitational moments (e.g. \cite{Guillot1999}) but the same H/He equation of state.  A state-of-the-art model for Jupiter in \cite{Hubbard2016} favors metal masses around 22 \me (but note that it uses a different EOS).}

\begin{table}[t!]
    \centering
    \begin{tabular}{|c|c|c|}
        \hline
         Source & Jupiter & Saturn \\
         \hline
         Guillot (1999) & 10-40 & 20-30 \\
         \hline
         This Work & 37 & 27 \\
         \hline
         $\pm 10\%$ M,R & $\pm20$ & $\pm5.5$ \\
         \hline
         $\pm 2$ Gigayears & $\pm1.1$ & $\pm.8$ \\
         \hline
    \end{tabular}
    \caption{\new{Inferred total heavy-element mass for Jupiter and Saturn, from \cite{Guillot1999} and this work.  For reference, we also show the uncertainties which would result if we had 10\% uncertainties in mass and radius, and separately for a 2 Gyr uncertainty in  age.  Note that the central values lie within the estimate from Guillot.}}
    \label{jsComparison}
\end{table}

\new{As we see from Table \ref{jsComparison}, our inferred metal masses fall within a plausible range for Jupiter and Saturn.  Furthermore, we show the errors resulting from uncertainties in mass and radius (10\% each) to demonstrate that these are the dominant sources of uncertainty in our study.  Also, we see that the error from age uncertainty for these somewhat old planets is not very significant.}
 
\section{Results} \label{Results}
We examined our results for connections between three quantities connected through the core-accretion model: the planetary mass $M$, its heavy element mass $M_z$, and the stellar metallicity [Fe/H].  We also considered the metal mass fraction of the planet \zpl and its ratio to that of the parent star \zpzs .

\subsection{Relation to Planet Mass}

\begin{figure}
    \centering
    \plotone{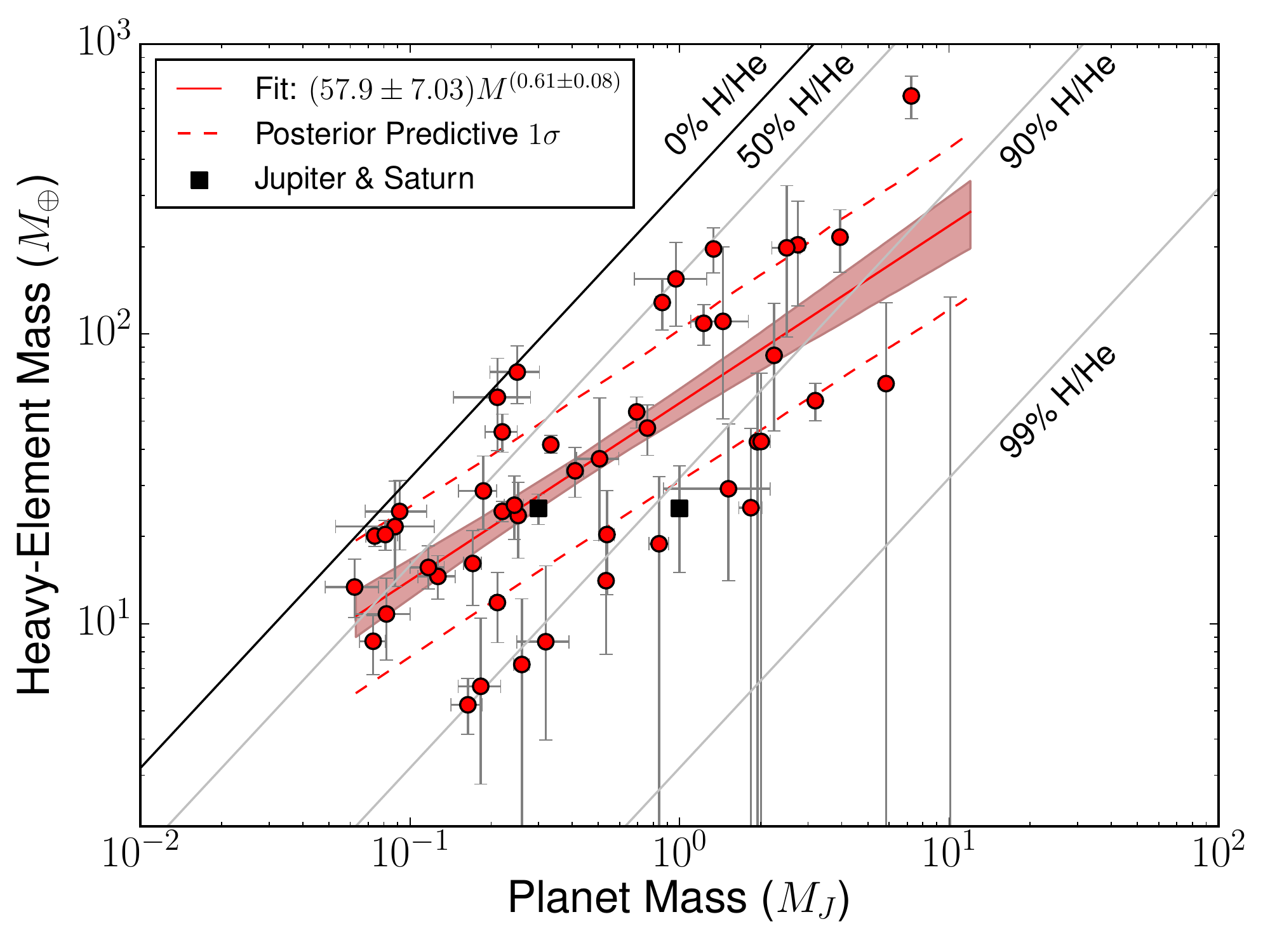}
    \caption{The heavy element masses of planets and their masses.  The lines of constant \zpl are shown at values of 1 (black), 0.5, 0.1, and .01 (Gray).  Distributions for points near \zpl = 1 tend to be strongly correlated (have well-defined \zpl values) but may have high mass uncertainties.  No models have a \zpl larger than one.  The distribution of fits (see \S \ref{Analysis} for discussion) is shown by a red median line with 1, 2, and 3 $\sigma$ contours.  Note Kepler-75b at 10.1 \mj\; which only has an upper limit.}
    \label{massMetal}
\end{figure}

There exists a clear correlation between planet mass and heavy element content.  We can see that as we move towards more massive planets, the total mass of heavy elements increases, but the bulk metallicity decreases (Figure \ref{massMetal}).  \new{Using Kendall's Tau with the mean values of metal mass and total planet mass, we measure a correlation of .4787 and a p-value of $2.07\times10^{-6}$, strongly supporting a correlation.}  This is consistent with a formation model where an initial heavy element core accretes predominantly, but not exclusively, H/He gas (e.g. \cite{Pollack1996}).  Indeed, \new{it appears likely that all of our sample planets have more than a few \me\; of heavy-elements and usually far more (though we cannot determine a minimum exactly), consistent} with both MF2011 and theoretical core-formation models \citep{Klahr2006}.  \new{A fit to the log of the data gives $M_z = (57.9\pm7.03) M^{(.61 \pm .08)}$, or roughly $M_z \propto \sqrt{M}$ and \zpl $\propto 1/\sqrt{M}$.}  Our parameter uncertainties exclude a flat line by a wide margin, but the distribution has a fair amount of spread around our fit. \new{The intrinsic spread was the factor $10^\sigma = 1.82\pm.09$ (because $\sigma$ was calculated on the log of the variables), which means that $1 \sigma$ of the data is within a factor of ~1.82 of the mean line.}  While some of this may be from observational uncertainty, it seems likely that other effects, such as the planet's migration history and the stochastic nature of planet formation, also play a role.  With this in mind, using planet mass alone to estimate the total heavy element mass appears accurate to a factor of a few.

\subsection{Effect of Stellar Metallicity}

\begin{figure}[t]
    \centering
    \plotone{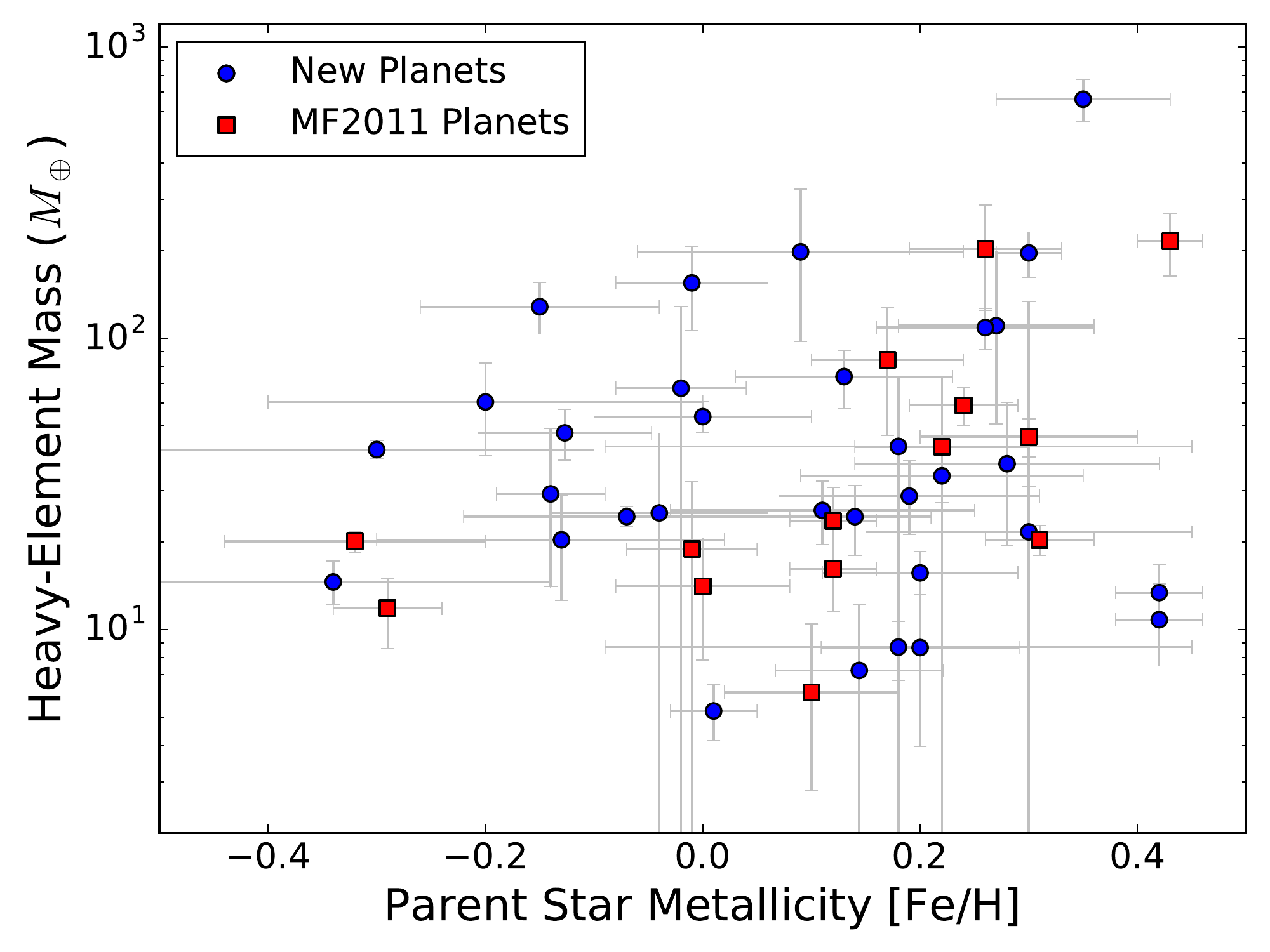}
    \caption{The heavy element masses of planets plotted against their parent star's metallicity.  Our results for the planets studied in MF2011 are in blue, and the remaining planets in our data set are in red.  A correlation appears for [Fe/H] in the blue points, but washes out with the new data.  Still, it appears that planets with high heavy element masses occur less frequently around low iron-metallicity stars.}
    \label{cmfe}
\end{figure}

The metallicity of a star directly impacts the metal content of its protoplanetary disk, increasing the speed and magnitude of heavy element accretion.  We examined our data for evidence of this connection. MF2011 observed a correlation for high metallicity parent stars between [Fe/H] and the heavy element masses of their planets (see also \cite{Guillot2006} and \cite{Burrows2007} for similar results from inflated planets).  If we constrain ourselves to the fourteen planets in MF2011, we see the same result.  However, the relation becomes somewhat murky for our full set of planets (see Figure \ref{cmfe}).  Applying Kendall's Tau to the most likely values of metal mass and [Fe/H], we measure a correlation of .08845 and a p-value of .3805, which indicates no correlation.  Some of the reason for this may lie with the high observational uncertainty in our values for stellar metallicity, but it is still difficult to believe that there is a direct power-law relationship.

\begin{figure}[t!]
    \centering
    \plotone{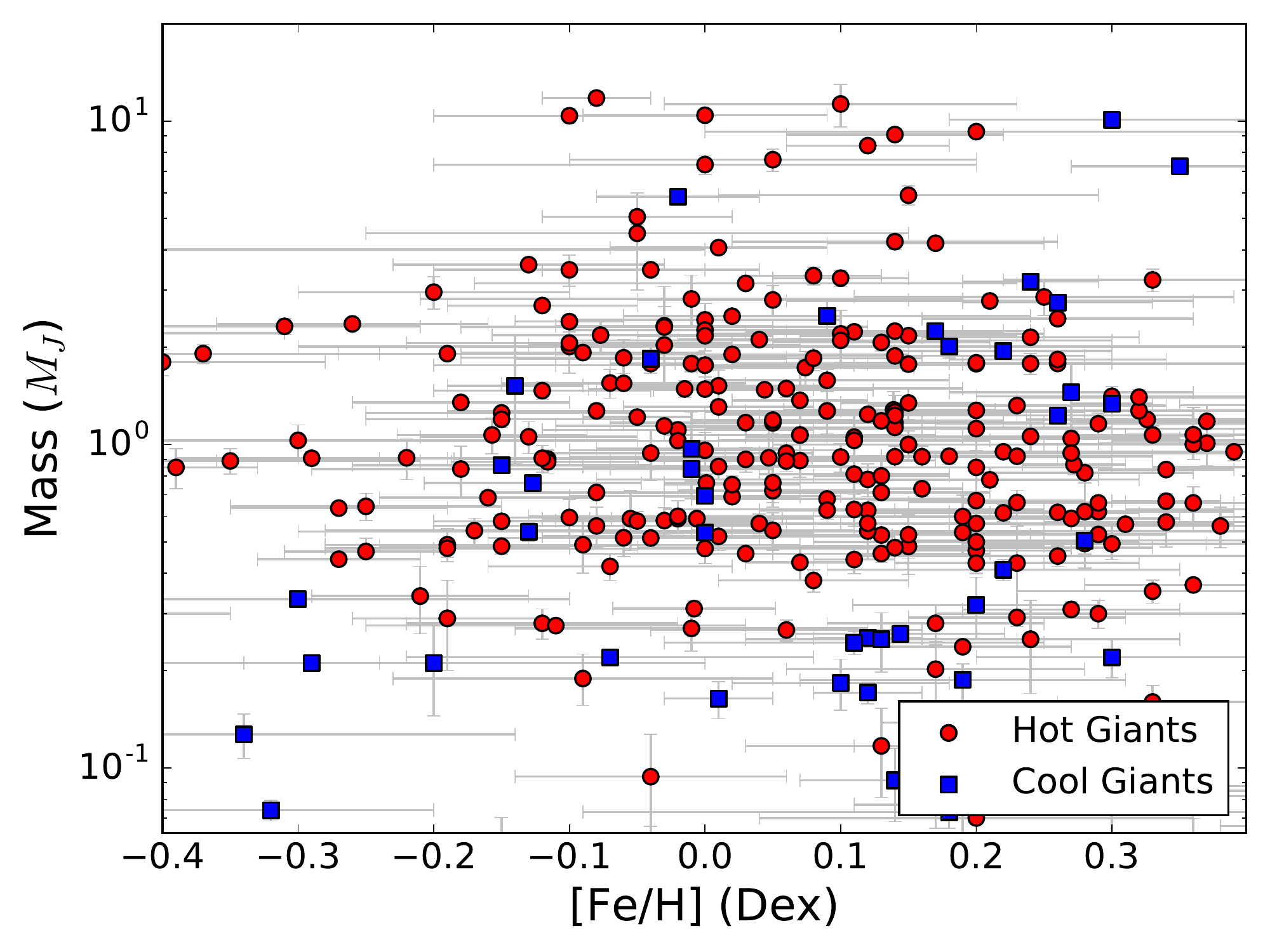}
    \caption{Planet mass plotted against parent star metallicity for transiting planets with RV masses.  Planets in our sample are in blue.  Planets in red were too strongly insolated to pass the flux cut (the inflated hot Jupiters).  Note the lack of planets around low-metallicity stars above about 1 \mj.  This, combined with the findings on Figure \ref{cmfe}, suggests that planets around low-metallicity stars are unable to generate the giant planets which typically have massive quantities of heavy elements.}
    \label{feMass}
\end{figure}

\new{Transit surveys should not be biased in stellar metallicity, so we can instead consider how the distribution of planet mass and metallicity vary as a function of stellar metallicity.}  Most of the planets with heavy element masses above 100\; \me\ orbit metal-rich stars; there is no clear pattern for planets with lower metal masses.  Considering the connection between planet mass and heavy element mass, we note that planets more massive than \new{2-3 Jupiters} are found far less often around low-metallicity stars (see Figure \ref{feMass}).  \new{Presumably, these trends are connected.}  This is similar to one of the findings of \cite{Fischer2005} in which the number of giant planets and the total detected planetary mass are correlated with stellar metallicity.  The population synthesis models in \cite{Mordasini2012} also observe and discuss an absence of very massive planets around metal-poor \new{stars.}  In the future, a more thorough look at this connection should take into account stellar metal abundances other than iron and put an emphasis on handling the high uncertainties in measurements of stellar metallicity.

\subsection{Metal Enrichment}

\begin{figure}[t!]
    \centering
    \plotone{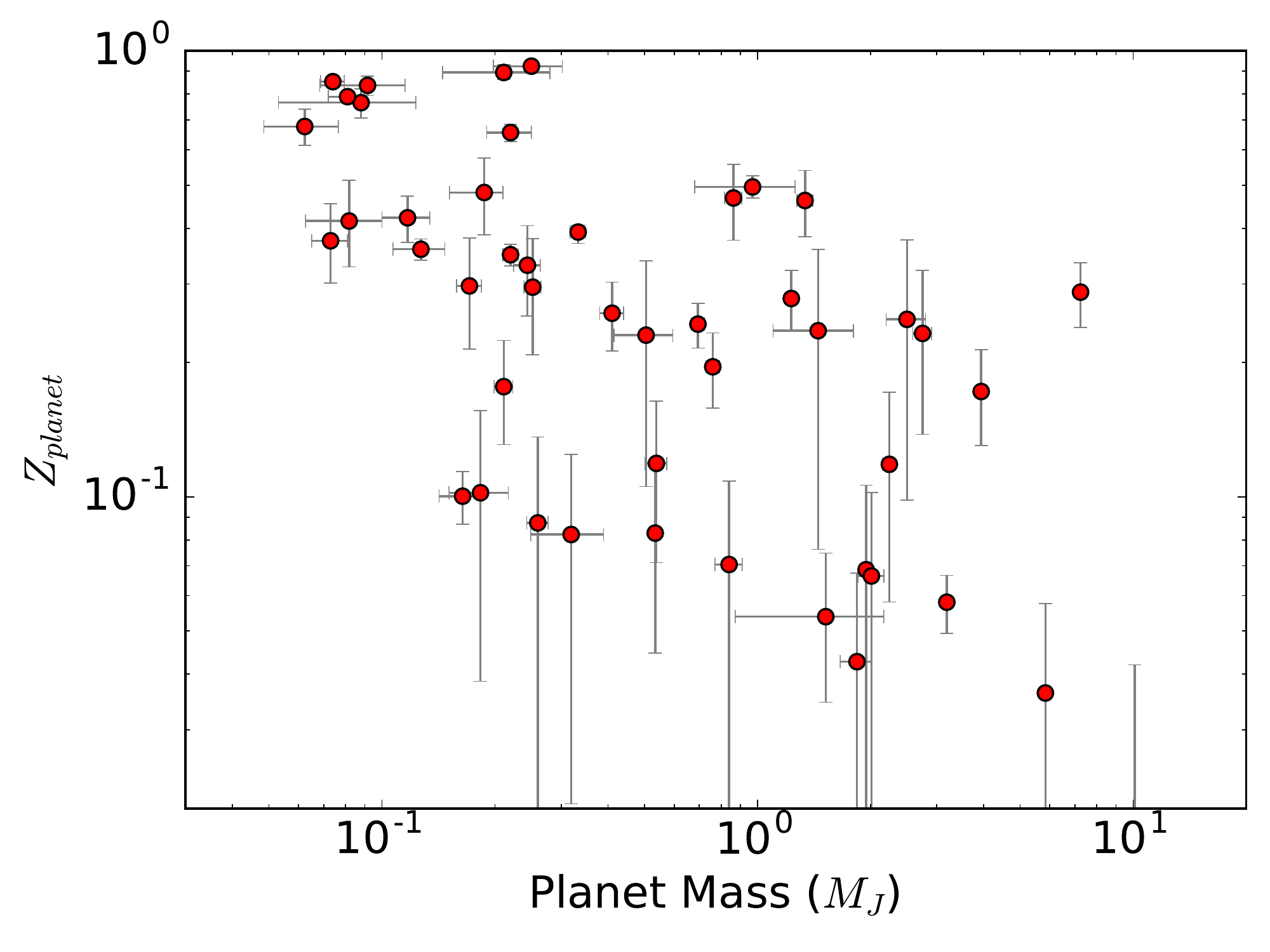}
    \caption{The heavy element fraction of planets as a function of mass.  We observe a downward trend with a fair amount of spread.  Compare especially with Figure \ref{massZZ}, which shows the same value relative to the parent star.}
    \label{massZpl}
\end{figure}

\begin{figure}[t!]
    \centering
    \plotone{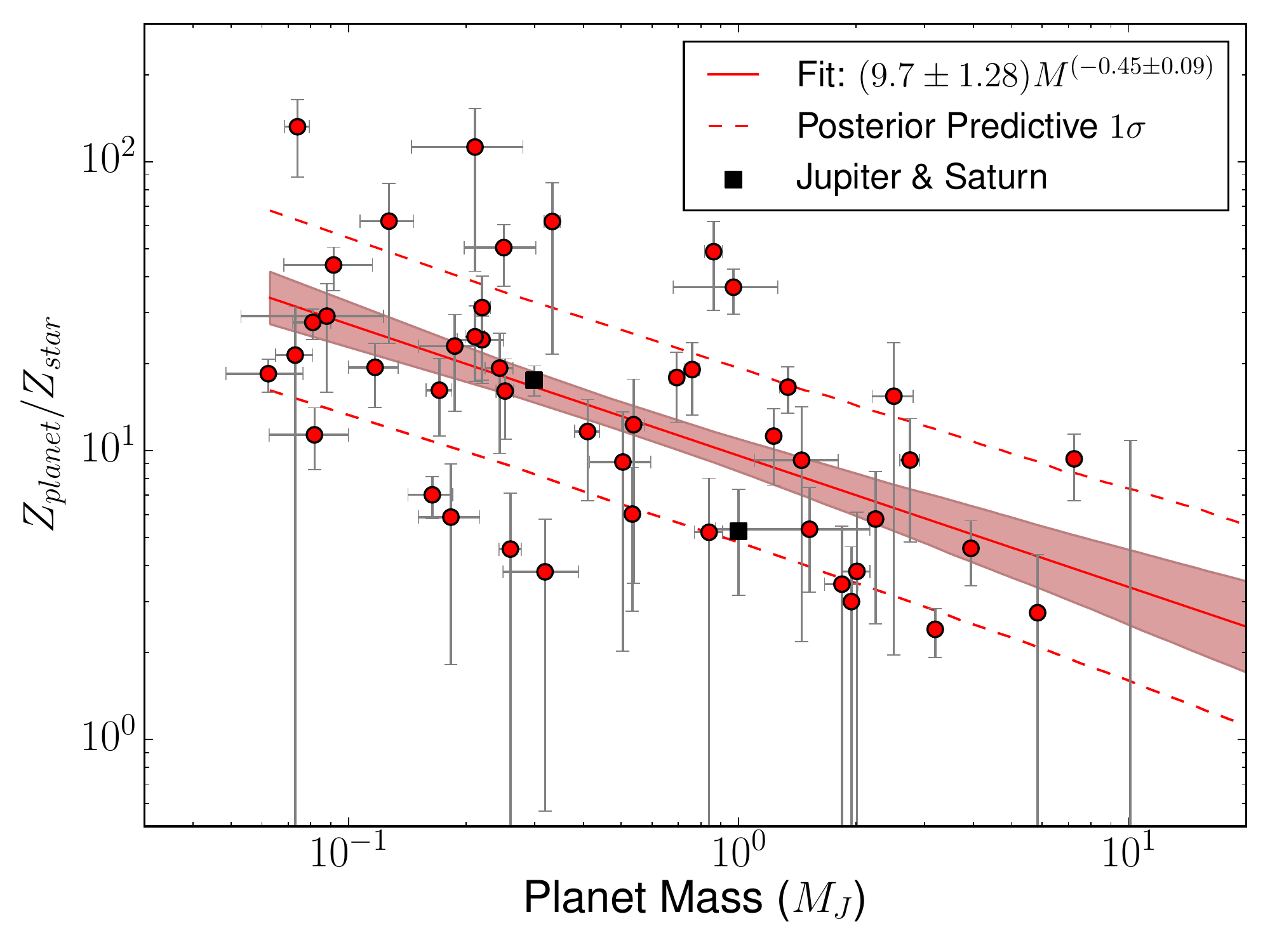}
    \caption{The heavy element enrichment of planets relative to their parent stars as a function of mass.  The line is our median fit to the distribution from bootstrapping, with 1, 2, and 3 $\sigma$ error contours.  Jupiter and Saturn are shown in blue, from \cite{Guillot1999}.  The pattern appears to be  stronger than considering \zpl\; alone against mass.}
    \label{massZZ}
\end{figure}

A negative correlation between a planet's metal enrichment relative to its parent star was suggested in MF2011 and found in subsequent population synthesis models \citep{Mordasini2014}, so we revisited the pattern with our larger sample.  We see a good relation in our data as well (Figure \ref{massZZ}), \new{and using Kendall's tau as before, we find a correlation of .4398 with a p-value of $1.3\times10^{-5}$.}  The exponent for the fit shown in Fig. \ref{massZZ} ($-.45 \pm .09$) differs somewhat from the planet formation models of \cite{Mordasini2014} (between -0.68 and -0.88).  The pattern appears to be stronger than if we considered only the planetary metal fraction \zpl alone (shown in Figure \ref{massZpl}).  This supports the notion that stellar metallicity still has some connection to planetary metallicity, even though we do not observe a power-law type of relation.  Jupiter and Saturn, shown in blue, fit nicely in the distribution.  Our results show that even fairly massive planets are enriched relative to their parent stars.  This is intriguing, because it suggests that (since their cores are probably not especially massive) the envelopes of these planets are strongly metal-enriched, a result which can be further tested through spectroscopy.  Note that we calculate our values of \zs\; by assuming that stellar metal scales with the measured iron metallicity [Fe/H], \new{using the simple approximation $Z_{star} = .014 \times 10^{[Fe/H]}$ (given our [Fe/H] uncertainties, a more advanced treatment would not be worthwhile).}  Considering other measurements of stellar metals in the future would be illuminating.  For instance, since oxygen in a dominant component of both water and rock, perhaps there exists a tighter correspondence between \zpl and the abundance of stellar oxygen, rather than with stellar iron.

\begin{figure}[t!]
    \centering
    \plotone{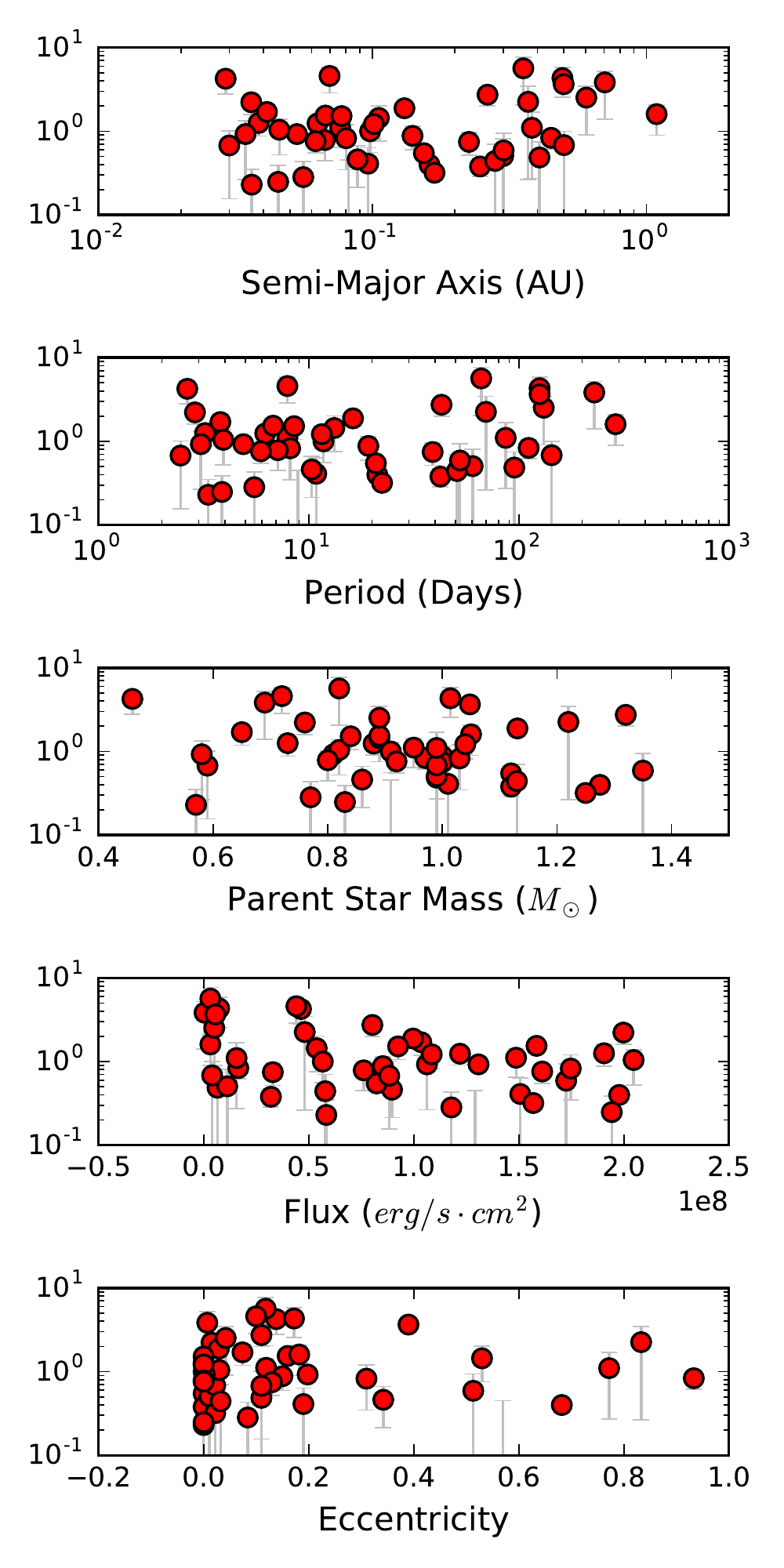}
    \caption{The relative residuals (calculated/fit) to the fit of \zpzs\, against mass, (Fig. \ref{massZZ}) plotted against the semi-major axis, period, parent star mass, flux, and eccentricity.  No relation is apparent.  The lack of a residual against flux implies that we have successfully cut out the inflated hot Jupiters; had we not, the high flux planets would be strong lower outliers.}
    \label{resids}
\end{figure}

\begin{figure}[t!]
    \centering
    \plotone{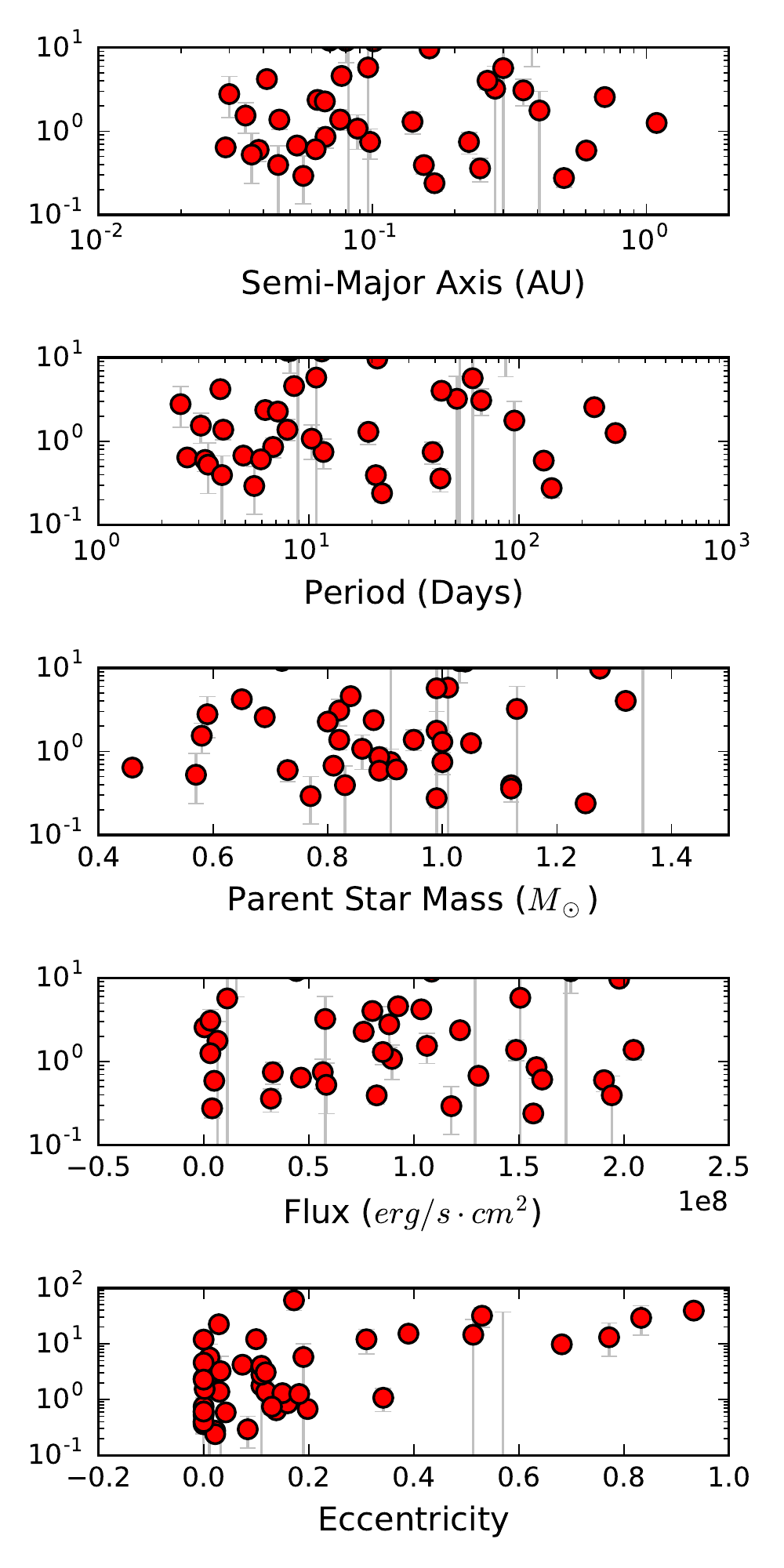}
    \caption{Same as Figure \ref{resids}, but for the fit on heavy-element mass against total mass seen in figure \ref{massMetal}.  The spread is somewhat higher in this case, but it exhibits a similar lack of correlation with flux.}
    \label{hResids}
\end{figure}

We also considered the possibility that orbital properties might relate to the metal content, perhaps as a proxy for the migration history.  We plot the residual from our mass vs. \zpzs\; fit against the semi-major axis, period, eccentricity, and parent star mass in Figure \ref{resids}.  No pattern is evident for any of these.  Given the number of planets and the size of our error-bars in our sample, we cannot rule out that any such patterns exist, but we do not observe them here.  We also considered the residual against the stellar flux.  Any giant planets with radius inflation which had made it into our sample would appear as strong outliers below the fit, since we would have mistaken inflation for lower heavy element masses.  Therefore, the lack of a pattern here suggests our flux cut is eliminating the inflated hot Jupiters as intended. 

\subsection{Heavy Element Masses in Massive Planets}
The extreme values for some of the heavy element masses are noteworthy.  HAT-P-20 b, the upper-right point in Figure \ref{massMetal}, contains over 600 \me \;of metal.  It is a 7.2 $M_J$ planet orbiting a metal-rich star ([Fe/H] = $.35\pm0.08$), so we expect it would be metal-rich.  Still, it is surprising that a planetary nebula can have so much metal to put into just one planet.  This is not a high estimate; our choice of a 10 \me \;core yields a value lower than if more of the metal were located in a core (see \S \ref{distribution}).  A core-dominated equivalent could have as much as 1000 \me \;of metal.  As such, it is much more plausible that the planet is envelope-dominated.  \new{The extreme metal content of HAT-P-20 b has been observed before \citep{Leconte2011} and is not unique (e.g. \cite{Cabrera2010}).  Such planets raise questions about how such extreme objects can form (see also \cite{Leconte2009} for a similar discussion for other massive giant planets).  These planets} would presumably have had to migrate through their system in such a way as to accumulate nearly all of the metal available in the disk.

However, in contrast with HAT-P-20b is Kepler-75b, at 10.1 \mj, our most massive planet in the sample.  It's metal-enrichment is significantly smaller than HAT-P-20b.  One could entertain the suggestion that HAT-P-20b formed via core-accretion, but that Kepler-75b is a low-mass brown dwarf that formed through a different mechanism.  At any rate, the future of determining whether a given object is a planet or low mass brown dwarf via characteristics like composition, rather than mass, which we advocate \cp[and see also][]{Chabrier2014} is promising.

\section{Interpretation}
\new{One might expect that core accretion produces giant planets with total metal masses of approximately $M_z = M_{\rm core} + Z_* M_{\rm env}$, where the core mass $M_{\rm core} \sim 10 M_\oplus$ depends on atmospheric opacity but is not known to depend strongly on final planet mass, a large fraction of disk solids are assumed to remain entrained with the envelope of mass $M_{\rm env}$ as it is accreted by the core, and the mass fraction of the disk in metals is assumed to match that of the star, $Z_*$.  Such a model would predict that a planet of total mass $M_p = M_{\rm core} + M_{\rm env}$ has a metallicity $Z_{pl} \equiv M_z/M_p$ that is related to that of the star by $Z_{pl}/Z_* = 1 + (M_{\rm core}/M_p)(1-Z_*)/Z_*$.  This expression falls off substantially more rapidly with planet mass than the $Z_{pl}/Z_* \approx 10 (M_p/M_J)^{-0.5}$ fit in Figure \ref{massZZ}.}

\new{This lack of good agreement is not surprising given that the above model does not predict the high metallicities of solar system giants.  These have long been interpreted as coming from late-stage accretion of additional planetesimal debris (e.g. \cite{mousis09}).  In keeping with this solar system intuition, we instead propose that the metallicity of a giant planet is determined by the isolation zone from which the planet can accrete solid material.  We assume that a majority of solids---which we treat interchangeably with metals in this initial investigation---eventually decouple from the disk gas and can then be accreted from the full gravitational zone of influence of the planet.  An object of mass $M$, accreting disk material with surface density $\Sigma_{a}$ at distance $r$ from a star of mass $M_*$, can accumulate a mass}

\begin{equation}\label{eqn-macc}
M_{a} = 2\pi r (2f_H R_H) \Sigma_{a} = 4\pi f_H \left(\frac{M}{3M_*}\right)^{1/3} \Sigma_{a} r^2
\end{equation}

\new{where $f_H \sim 3.5$ is the approximate number of Hill radii $R_H = r(M/3M_*)^{1/3}$ from which accretion is possible as long as the orbital eccentricity of accreted material is initially less than $(M/3M_*)^{1/3}$ \citep{Lissauer93}.  Thus, under our assumption that solids have decoupled from the gas, a planet of mass $M_p$ can accrete a total solid mass of}

\begin{equation}\label{eqn-msiso}
M_z \approx 4\pi f_H f_e Z_* \left(\frac{M_p}{3M_*}\right)^{1/3} \Sigma r^2 \;\;,
\end{equation}

\new{where $\Sigma$ is the total surface density of the disk and $f_e Z_* \Sigma$ is the surface density in solids.   The parameter $f_e$ allows for an enhancement in the metal mass fraction of the disk compared to the solar value, $Z_*$, for example due to radial drift of solid planetesimals through the gas nebula.  No enhancement corresponds to $f_e = 1$.  We note that Equation \ref{eqn-msiso} applies independent of the solid mass fraction of the planet because $M_p$ is taken to be the final observed planet mass, including any accreted solids.  }

\new{For comparison, the standard isolation mass of a planet forming in a disk with total surface density $\Sigma$ is}
\begin{equation}\label{eqn-miso}
M_{iso} = \left[ 4\pi f_H (3M_*)^{-1/3} \Sigma r^2 \right]^{3/2}
\end{equation}
which may be calculated using Equation \ref{eqn-macc} with $M_{a}=M=M_{iso}$ and $\Sigma_{a} = \Sigma$.  
\new{Recalling that $Z_{pl} = M_z/M_p$, Equations \ref{eqn-msiso} and \ref{eqn-miso} combine to yield}
\begin{equation}\label{eqn-zrat}
\frac{Z_{pl}}{Z_*} = f_e \left(\frac{M_p}{M_{iso}}\right)^{-2/3}
\end{equation}
\new{For $f_e = 1$, when $M_p = M_{iso}$, the total mass of the planet equals the total disk mass in its isolation region and $Z_{pl}/Z_* = 1$, as, for example, would happen if an isolation-mass planet formed by accumulating all material within its isolation zone.   For $M_p < M_{iso}$, the planet's metallicity exceeds the metallicity of the star because the planet has not been able to accrete all isolation-zone gas but we assume that it is able to accrete all of the region's solids.}

\begin{figure}[h]
\begin{center}
\plotone{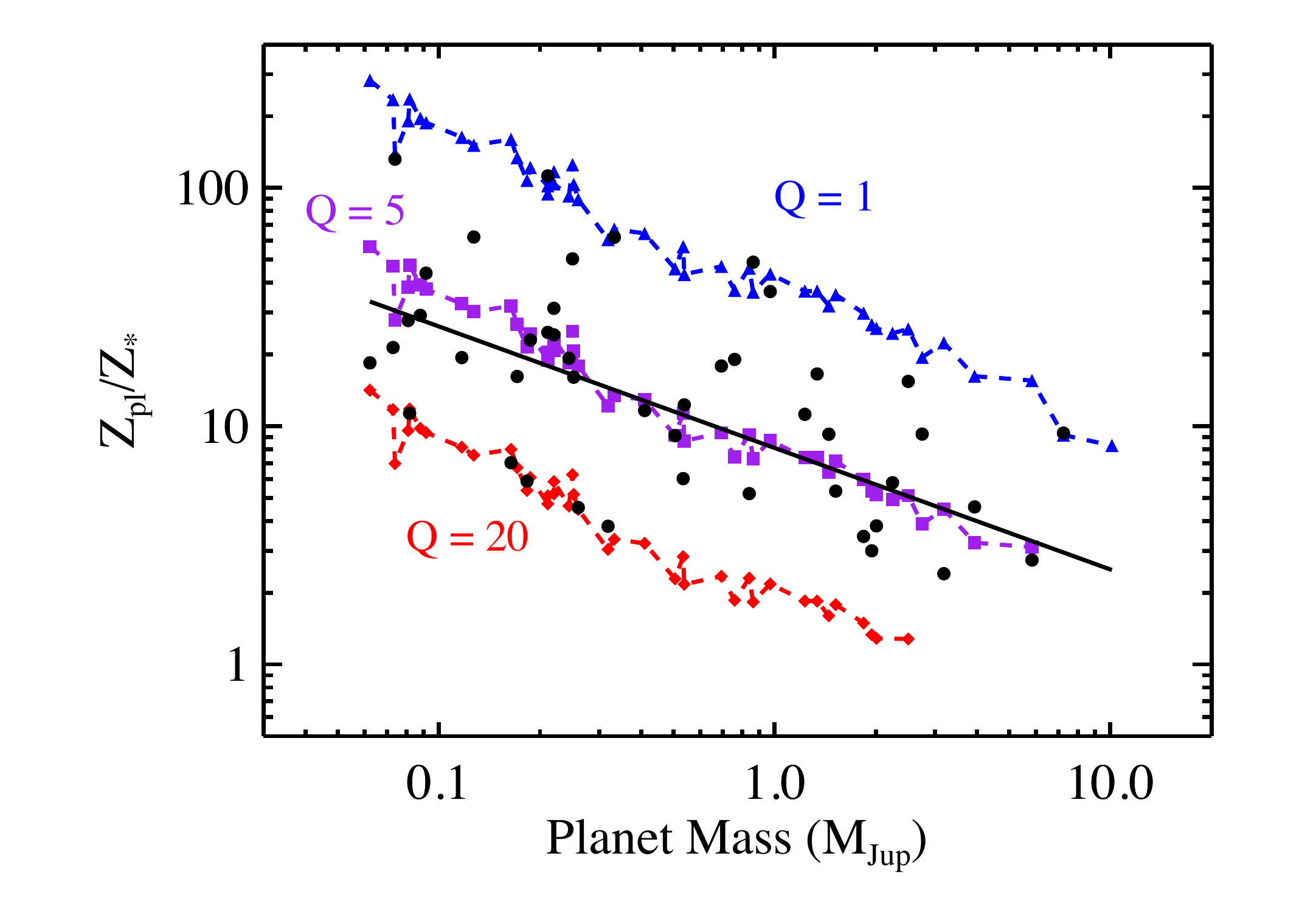}
\caption{Metallicity ratio calculated using Equation \ref{eqn-zrat2} with $f_e = 1$ using $M_p$ and $M_*$ for each of the planets in Figure Y.  We set $H/r = 0.04$ and plot curves for $Q = 1$ (blue triangles), 5 (purple squares), and 20 (red diamonds). The best fit line (in Figure \ref{massZZ}, solid black) for the data (black circles) is displayed for reference.}\label{fig-zmodel}
\end{center}
\end{figure}

\begin{figure}[h]
\begin{center}
\plotone{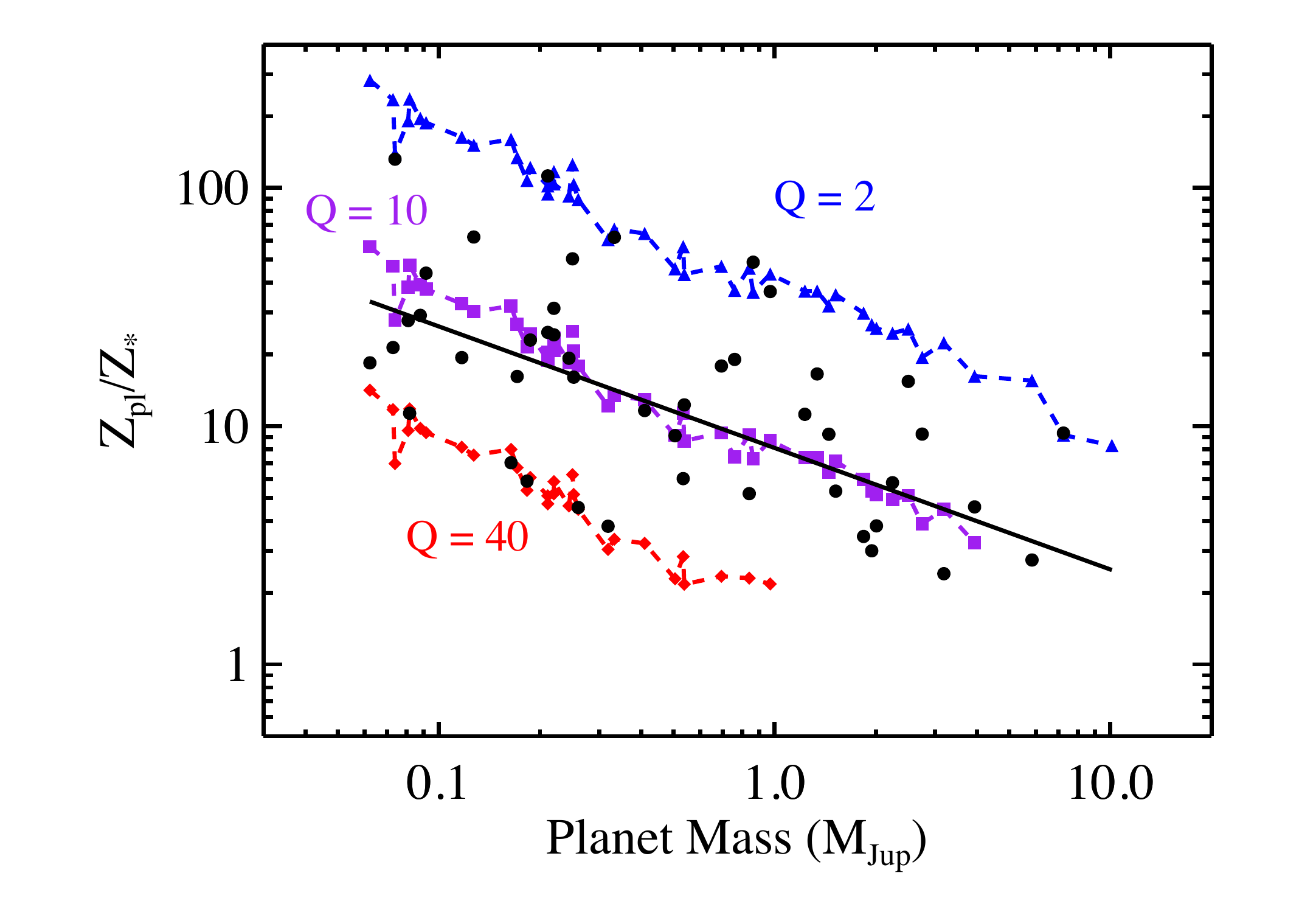}
\caption{Same as Figure \ref{fig-zmodel} but with $f_e = 2$.}\label{fig-zmodel2}
\end{center}
\end{figure}

\new{Equation \ref{eqn-zrat} encapsulates the physics of our model, but to compare with observed planets, we re-express this result in terms more easily related to the expected population of protoplanetary disks from which planets form---Toomre's $Q$ parameter \citep{Safronov60,Toomre64} and the disk aspect ratio $H/r$.  In a Keplerian disk, $Q = c_s\Omega/(\pi G\Sigma) = (H/r)(\pi r^2 \Sigma/M_*)^{-1}$, where $c_s$ is the isothermal sound speed, $\Omega$ is the orbital angular velocity, $H = c_s/\Omega$ is the disk scale height, and $G$ is the gravitational constant.  Thus,}
\begin{equation}\label{eqn-zrat2}
\frac{Z_{pl}}{Z_*} \approx 3 f_H f_e \frac{H}{r} Q^{-1} \left(\frac{M_p}{M_*}\right)^{-2/3}
\end{equation}
\new{Figure \ref{fig-zmodel} replots the values and best fit line of Figure \ref{massZZ}, overlaying $Z_{pl}/Z_*$ calculated using Equation \ref{eqn-zrat2} with $f_e=1$ for each planet's $M_p$ and $M_*$.  We use a fiducial value of $H/r = 0.04$ (corresponding to e.g., 2 AU at 200 K) and plot curves for $Q=1$, 5, and 20.  The $Q=5$ curve provides a good match for the best fit line, while $Q=1$ and $Q=20$ bound the remaining data points.  We omit data points corresponding to planet masses in excess of the total local disk mass, $M_p > \pi r^2 \Sigma = (H/r)Q^{-1} M_*$, which removes the portion of the $Q=20$ curve corresponding to the highest planet masses.  For $f_e>1$, the values of $Q$ plotted in Figure \ref{fig-zmodel} should be multiplied by $f_e$.  Figure \ref{fig-zmodel2} provides an example for $f_e = 2$, demonstrating that for modest enhancements in the solid to gas ratio in the disk, limits on the total disk mass could preclude formation of the most massive planets in the sample in the highest $Q$ disks, potentially explaining the lack of points in the bottom-right portion of the plot.}
\new{Values of $Q<1$ imply gravitational instability and cannot be maintained for extended periods in a disk, so the fact that our modeled $Z_{pl}/Z_*$ does not require $Q<1$ for any value of $f_e$ is encouraging.  For reference, at Jupiter's location $r=5$AU in the minimum mass solar nebula $\Sigma = 2\times 10^3$g cm$^{-2}$ $(a/{\rm AU})^{-3/2}$ \citep{Hayashi1981}, $Q \approx 25$.}

\new{Why might a planet remain smaller than its isolation mass?  For our model, the reason is irrelevant.  The planet could have accreted its envelope at a different location in the disk at an earlier time when solids were unable to decouple from the gas.  It could even be a remnant fragment formed by gravitational instability.  Whatever its formation process, our model only asserts that it accretes most of its metals at some point concurrent with or after it has accumulated most of the gas it will accrete from the nebula.  Hence, its metallicity is determined by the mass in solids contained within its gravitational feeding zone.  However, for a plausible scenario, we may again appeal to studies of Jupiter formation, which suggest that the planet's final mass is determined by the mass at which the planet truncates gas accretion by opening a large gap in the disk (e.g. \cite{Lissauer09}).}

\begin{figure}[h]
\begin{center}
\plotone{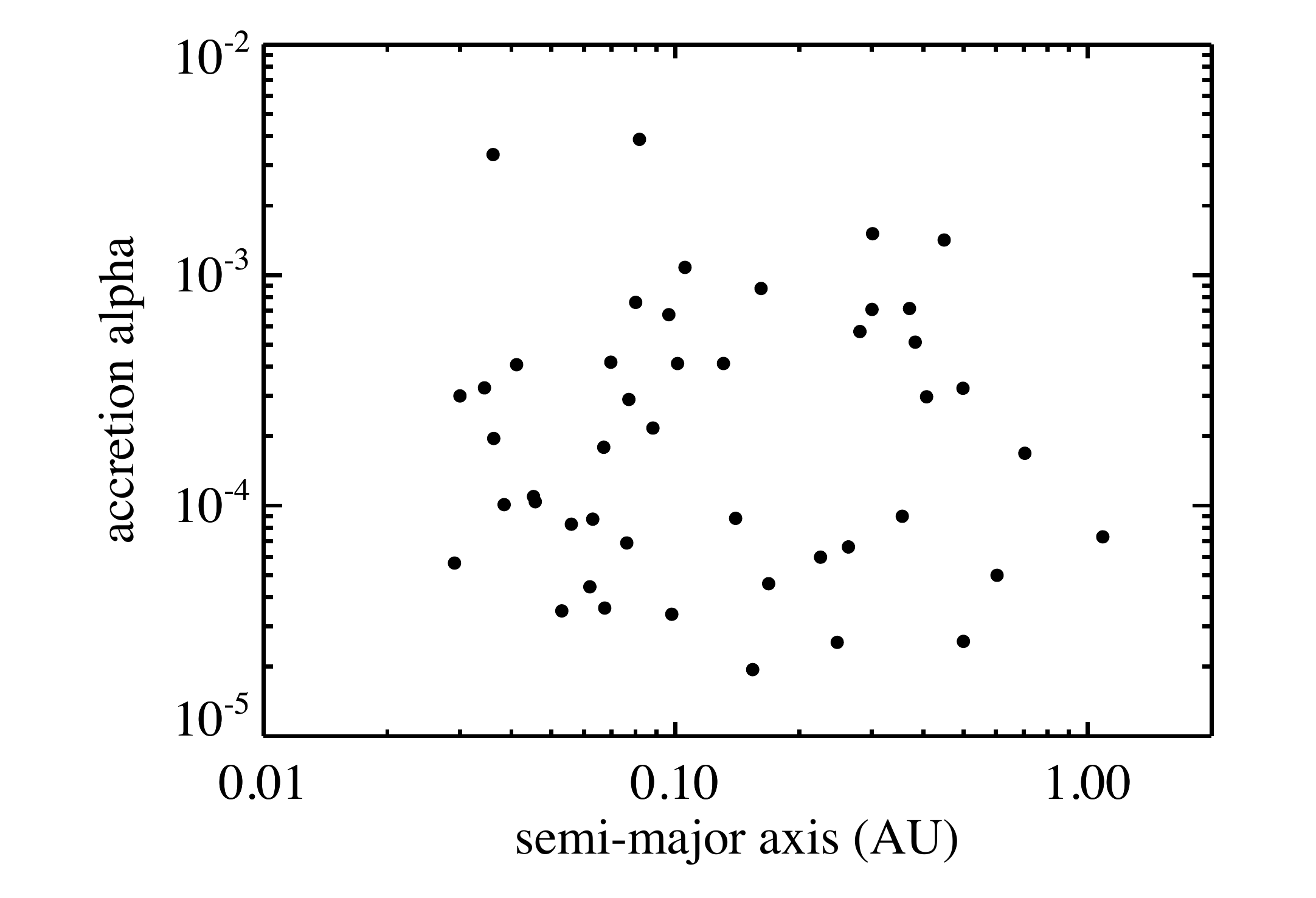}
\caption{Implied values of the disk viscosity parameter $\alpha$, calculated using Equation \ref{eqn-alpha}, for the scenario that the total planet masses observed in our sample are limited to less than the isolation mass at their formation locations due to gap starvation.  These values span a reasonable theoretical range for protoplanetary disks.}\label{fig-aalpha}
\end{center}
\end{figure}

\new{Gap opening and the planetary starvation that results remains a topic of continuing research (e.g. \cite{Crida2006},\cite{Fung2014},\cite{Duffell2015}).  Here we simply note that the planet masses observed in our sample are consistent with a classic theory of gap starvation.  Tidal torques opposed by viscous accretion yield a gap width }
\begin{equation}
\Delta = \left(\frac{f_g}{\pi \alpha} \frac{M_p}{M_*} \frac{r^2}{H^2}\right)^{1/3} R_H \;\;,
\end{equation}
\new{where $f_g \approx 0.23$ is a geometric factor \citep{Lin93}.  Assuming that gas accretion through the gap becomes inefficient for $\Delta/R_H = f_S \sim 5$ \citep{Lissauer09,Kratter10}, a planet mass $M_p$ set by gap truncation implies the disk viscosity parameter (\cite{Shakura1973}, \cite{Armitage2011})
}\begin{equation}\label{eqn-alpha}
\alpha = \frac{f_g}{\pi f_S^3} \frac{M_p}{M_*} \left(\frac{H}{r}\right)^{-2} \;\;.
\end{equation}
\new{These values---displayed in Figure \ref{fig-aalpha} for the same planets plotted in Figure \ref{fig-zmodel}---span a reasonable theoretical range for protoplanetary disks, particularly in dead zones, where most giant planets are thought to form (e.g. \cite{Turner2014}, \cite{Bai2016}).}

\new{We note that Equations \ref{eqn-zrat2} and \ref{eqn-alpha} do not depend directly on the planet's distance from its star, $r$.  Instead, they are functions of $H/r$, which depends weakly on $r$ for typical disks, and of $Q$, which may be thought of as a parametrization of the disk mass.  Hence, our model applies regardless of whether the observed planets have migrated from their formation locations.  For relatively low-mass disks, such migration is in fact required.  For the minimum mass solar nebula used above, for example, $M_{iso} = 0.6 M_J (r/{\rm AU})^{3/4}$.  Giant planets separated by less than 1AU from their host stars most likely did migrate from more distant formation locations \citep[e.g.][]{Dawson13}.}

\section{Discussion and Conclusions}
There is a strong connection between planet mass and metal content.  From a sample of 47 transiting gas giants, we find that the heavy-element mass increases as $\sqrt{M}$, so \zpl decreases as $1/\sqrt{M}$.  We also see that our planets are consistently enriched relative to their parent stars, and that they \new{likely all have more than a few \me\; of heavy-elements}.  These results all support the core-accretion model of planet formation and the previous results from MF2011 that metal-enrichment is a defining characteristic of giant planets.  \new{We have also shown that our results for \zpzs are comfortably consistent with a simple planet formation model using plausible values for disk parameters.  Our results were not consistent with a more naive model of formation in which a fixed-mass core of heavy-elements directly accretes parent-star composition nebular material.}

This work suggests that spectroscopy of the atmospheres of gas giants should also yield metal-enrichments compared to parent star abundances \cp{Fortney2008}, as is seen in the solar system.  We suggest that this growing group of $< 1000$ K planets should be a sample of great interest for atmospheric observations with the \emph{James Webb Space Telescope}.  Since the bulk metallicity of the cool planets can be determined, atmospheric studies to determine metal-enrichments can be validated.  For most of our planets more massive than Saturn, the heavy elements values are high enough that most metals are in the envelope, rather than the core (See Figure \ref{massMetal}), so our \zpl\ values are upper limits on $Z_{\rm atmosphere}$.  Atmospheric observations to retrieve the mixing ratios of abundant molecules like H$_2$O, CO, CO$_2$, and CH$_4$ could show if most heavy elements are found within giant planet envelopes or within cores.  In comparison, for hot Jupiters $> 1000$ K we cannot estimate with confidence the planetary bulk metal enrichment since the radius inflation power is unknown.

Some connection between planetary heavy element mass and stellar metallicity is suggested by our work, but the correlation is not strong.  A fruitful area of future investigation will be in analyzing stellar abundances other than iron.  Learning which stellar metals most strongly correlate with planetary bulk metallicity (for instance, Fe, Mg, Si, Ni, O, C) would hint at the composition of planetary heavy elements and could provide insights into the planet formation process.  With the continued success of ground-based transit surveys, along with K2 \citep{Howell2014}, and the 2017 launch of \emph{TESS}, we expect to see many more ``cooler" gas giant planets amenable to this type of analysis, which will continue to provide an excellent opportunity for further exploring these relations.  With this continuing work on bulk metal-enrichment of planets, and the spectroscopy of planetary atmospheres, the move towards understanding planet formation in the mass / semi-major axes / composition planes should be extremely fruitful.

\begin{deluxetable}{llllllllll}\label{resultTable}
\rotate
\tabletypesize{\tiny}
\tablehead{
\colhead{\#} & \colhead{Name} & \colhead{Mass (\mj)}& \colhead{Radius (\rj)}& \colhead{Age (Gyr)}& \colhead{Flux ($\frac{erg}{\text{cm}^2\cdot s}$)} & \colhead{[Fe/H] (Dex)} & \colhead{Metal (\me)} & \colhead{$Z_p$} & \colhead{$Z_p/Z_s$}
}
\startdata
1& CoRoT-8 b  &  $    0.22\pm0.03$  &  $    0.57\pm0.02$  &  $     0.1-     3.0$
 &  ${1.22\cdot 10^{8}}$
 &  $    0.30\pm0.10$  &  $   45.91_{-6.83}^{+6.99}$  &  $    0.66\pm0.03$  &  $   24.18_{-6.67}^{+5.00}$ \\
2$^\dagger$& CoRoT-9 b  &  $    0.84\pm0.07$  &  $    1.05\pm0.04$  &  $     1.0-     8.0$
 &  ${6.59\cdot 10^{6}}$
 &  $   -0.01\pm0.06$  &  $   18.87_{-18.77}^{+13.30}$  &  $    0.07_{-0.07}^{+0.04}$  &  $    5.21_{-5.21}^{+2.81}$ \\
3& CoRoT-10 b  &  $    2.75\pm0.16$  &  $    0.97\pm0.07$  &  $     0.1-     3.0$
 &  ${5.38\cdot 10^{7}}$
 &  $    0.26\pm0.07$  &  $  202.98_{-78.08}^{+84.09}$  &  $    0.23\pm0.09$  &  $    9.26_{-4.44}^{+3.63}$ \\
4& CoRoT-24 c  &  $    0.09\pm0.04$  &  $    0.44\pm0.04$  &  $     8.0-    14.0$
 &  ${5.67\cdot 10^{7}}$
 &  $    0.30\pm0.15$  &  $   21.64_{-8.17}^{+9.38}$  &  $    0.77\pm0.06$  &  $   29.18_{-13.28}^{+8.55}$ \\
5& GJ 436 b  &  $    0.07\pm0.01$  &  $    0.37\pm0.02$  &  $     0.5-    10.0$
 &  ${4.63\cdot 10^{7}}$
 &  $   -0.32\pm0.12$  &  $   20.07_{-1.62}^{+1.68}$  &  $    0.85\pm0.03$  &  $  132.02_{-43.68}^{+31.66}$ \\
6& HAT-P-11 b  &  $    0.08\pm0.01$  &  $    0.42\pm0.01$  &  $     2.4-    12.4$
 &  ${1.31\cdot 10^{8}}$
 &  $    0.31\pm0.05$  &  $   20.31_{-2.36}^{+2.45}$  &  $    0.79\pm0.02$  &  $   27.76_{-3.56}^{+3.09}$ \\
7& HAT-P-12 b  &  $    0.21\pm0.01$  &  $    0.96_{-0.02}^{+0.03}$  &  $     0.5-     4.5$
 &  ${1.91\cdot 10^{8}}$
 &  $   -0.29\pm0.05$  &  $   11.83_{-3.23}^{+3.17}$  &  $    0.18\pm0.05$  &  $   24.75_{-7.43}^{+6.92}$ \\
8$^\dagger$& HAT-P-15 b  &  $    1.95\pm0.07$  &  $    1.07\pm0.04$  &  $     5.2-     9.3$
 &  ${1.51\cdot 10^{8}}$
 &  $    0.22\pm0.08$  &  $   42.45_{-42.35}^{+30.67}$  &  $    0.07_{-0.07}^{+0.04}$  &  $    3.00_{-3.00}^{+1.65}$ \\
9& HAT-P-17 b  &  $    0.53\pm0.02$  &  $    1.01\pm0.03$  &  $     4.5-    11.1$
 &  ${8.97\cdot 10^{7}}$
 &  $    0.00\pm0.08$  &  $   14.08_{-6.24}^{+6.56}$  &  $    0.08\pm0.04$  &  $    6.02_{-3.25}^{+2.72}$ \\
10& HAT-P-18 b  &  $    0.18\pm0.03$  &  $    0.95\pm0.04$  &  $     6.0-    16.8$
 &  ${1.18\cdot 10^{8}}$
 &  $    0.10\pm0.08$  &  $    6.08_{-3.29}^{+4.35}$  &  $    0.10_{-0.06}^{+0.05}$  &  $    5.88_{-4.06}^{+3.11}$ \\
11& HAT-P-20 b  &  $    7.25\pm0.19$  &  $    0.87\pm0.03$  &  $     2.9-    12.4$
 &  ${2.00\cdot 10^{8}}$
 &  $    0.35\pm0.08$  &  $  662.48_{-109.11}^{+111.85}$  &  $    0.29\pm0.05$  &  $    9.35_{-2.66}^{+2.06}$ \\
12& HAT-P-54 b  &  $    0.76\pm0.03$  &  $    0.94\pm0.03$  &  $     1.8-     8.2$
 &  ${1.04\cdot 10^{8}}$
 &  $   -0.13\pm0.08$  &  $   47.33_{-9.22}^{+9.60}$  &  $    0.20\pm0.04$  &  $   19.06_{-5.80}^{+4.54}$ \\
13& HATS-6 b  &  $    0.32\pm0.07$  &  $    1.00\pm0.02$  &  $     0.1-    13.7$
 &  ${5.84\cdot 10^{7}}$
 &  $    0.20\pm0.09$  &  $    8.67_{-4.70}^{+7.17}$  &  $    0.08_{-0.06}^{+0.04}$  &  $    3.80_{-3.24}^{+1.98}$ \\
14& HATS-17 b  &  $    1.34\pm0.07$  &  $    0.78\pm0.06$  &  $     0.8-     3.4$
 &  ${9.97\cdot 10^{7}}$
 &  $    0.30\pm0.03$  &  $  196.42_{-34.17}^{+35.50}$  &  $    0.46\pm0.08$  &  $   16.56_{-3.12}^{+2.93}$ \\
15& HD 17156 b  &  $    3.19\pm0.03$  &  $    1.09\pm0.01$  &  $     2.4-     3.8$
 &  ${1.98\cdot 10^{8}}$
 &  $    0.24\pm0.05$  &  $   58.84_{-8.75}^{+8.81}$  &  $    0.06\pm0.01$  &  $    2.40_{-0.49}^{+0.43}$ \\
16& HD 80606 b  &  $    3.94\pm0.11$  &  $    0.98\pm0.03$  &  $     1.7-     7.6$
 &  ${1.65\cdot 10^{7}}$
 &  $    0.43\pm0.03$  &  $  215.69_{-52.23}^{+52.63}$  &  $    0.17\pm0.04$  &  $    4.58_{-1.19}^{+1.13}$ \\
17& K2-19 b  &  $    0.19_{-0.04}^{+0.02}$  &  $    0.67\pm0.07$  &  $     8.0-    14.0$
 &  ${1.49\cdot 10^{8}}$
 &  $    0.19\pm0.12$  &  $   28.72_{-7.56}^{+9.21}$  &  $    0.48\pm0.09$  &  $   22.99_{-9.31}^{+6.59}$ \\
18& K2-24 b  &  $    0.06\pm0.01$  &  $    0.52\pm0.05$  &  $     1.3-     8.9$
 &  ${8.24\cdot 10^{7}}$
 &  $    0.42\pm0.04$  &  $   13.38_{-2.86}^{+3.30}$  &  $    0.68\pm0.06$  &  $   18.44_{-2.56}^{+2.27}$ \\
19& K2-24 c  &  $    0.08\pm0.02$  &  $    0.72\pm0.07$  &  $     1.3-     8.9$
 &  ${3.20\cdot 10^{7}}$
 &  $    0.42\pm0.04$  &  $   10.81_{-3.32}^{+3.55}$  &  $    0.42_{-0.09}^{+0.10}$  &  $   11.33\pm2.73$ \\
20& K2-27 b  &  $    0.09\pm0.02$  &  $    0.40\pm0.03$  &  $     0.5-    10.0$
 &  ${1.58\cdot 10^{8}}$
 &  $    0.14\pm0.07$  &  $   24.40_{-6.42}^{+6.86}$  &  $    0.84\pm0.04$  &  $   43.88_{-8.19}^{+6.65}$ \\
21& Kepler-9 b  &  $    0.25\pm0.01$  &  $    0.84\pm0.07$  &  $     2.0-     4.0$
 &  ${8.53\cdot 10^{7}}$
 &  $    0.12\pm0.04$  &  $   23.63_{-6.80}^{+7.15}$  &  $    0.30_{-0.09}^{+0.08}$  &  $   16.06_{-5.11}^{+4.69}$ \\
22& Kepler-9 c  &  $    0.17\pm0.01$  &  $    0.82\pm0.07$  &  $     2.0-     4.0$
 &  ${3.29\cdot 10^{7}}$
 &  $    0.12\pm0.04$  &  $   16.14_{-4.58}^{+4.81}$  &  $    0.30\pm0.08$  &  $   16.17_{-4.95}^{+4.61}$ \\
23$^*$& Kepler-16 b  &  $    0.33\pm0.02$  &  $    0.75\pm0.00$  &  $     0.5-    10.0$
 &  ${4.19\cdot 10^{5}}$
 &  $   -0.30\pm0.20$  &  $   41.49_{-2.78}^{+3.11}$  &  $    0.39_{-0.02}^{+0.01}$  &  $   62.05_{-40.47}^{+22.31}$ \\
24$^\dagger$& Kepler-30 c  &  $    2.01\pm0.16$  &  $    1.10\pm0.04$  &  $     0.2-     3.8$
 &  ${1.12\cdot 10^{7}}$
 &  $    0.18\pm0.27$  &  $   42.54_{-42.44}^{+30.62}$  &  $    0.07_{-0.07}^{+0.04}$  &  $    3.81_{-3.81}^{+2.30}$ \\
25& Kepler-30 d  &  $    0.07\pm0.01$  &  $    0.79\pm0.04$  &  $     0.2-     3.8$
 &  ${4.05\cdot 10^{6}}$
 &  $    0.18\pm0.27$  &  $    8.70_{-2.02}^{+1.95}$  &  $    0.38_{-0.07}^{+0.08}$  &  $   21.39_{-21.53}^{+10.03}$ \\
26$^*$& Kepler-34 b  &  $    0.22\pm0.01$  &  $    0.76\pm0.01$  &  $     5.0-     6.0$
 &  ${3.21\cdot 10^{6}}$
 &  $   -0.07\pm0.15$  &  $   24.42_{-1.90}^{+1.97}$  &  $    0.35\pm0.02$  &  $   31.24_{-14.14}^{+8.98}$ \\
27$^*$& Kepler-35 b  &  $    0.13\pm0.02$  &  $    0.73\pm0.01$  &  $     8.0-    12.    $
 &  ${5.06\cdot 10^6}$
 &  $   -0.34\pm0.20$  &  $   14.56_{-2.40}^{+2.62}$  &  $    0.36\pm0.02$  &  $   62.22_{-38.71}^{+21.89}$ \\
28& Kepler-45 b  &  $    0.51\pm0.09$  &  $    0.96\pm0.11$  &  $     0.4-     1.5$
 &  ${8.83\cdot 10^{7}}$
 &  $    0.28\pm0.14$  &  $   37.08_{-17.71}^{+23.05}$  &  $    0.23_{-0.12}^{+0.11}$  &  $    9.12_{-7.11}^{+4.46}$ \\
29$^\dagger$& Kepler-75 b  &  $   10.10\pm0.40$  &  $    1.05\pm0.03$  &  $     3.4-     9.7$
 &  ${1.29\cdot 10^{8}}$
 &  $    0.30\pm0.12$  &  $    0.00_{-0.00}^{+133.99}$  &  $    0.00_{-0.00}^{+0.04}$  &  $    0.00_{-0.00}^{+10.84}$ \\
30& Kepler-89 d  &  $    0.16\pm0.02$  &  $    0.98\pm0.01$  &  $     3.7-     4.2$
 &  ${1.57\cdot 10^{8}}$
 &  $    0.01\pm0.04$  &  $    5.25_{-1.10}^{+1.23}$  &  $    0.10\pm0.01$  &  $    7.02_{-1.21}^{+1.10}$ \\
31$^\dagger$& Kepler-117 c  &  $    1.84\pm0.18$  &  $    1.10\pm0.04$  &  $     3.9-     6.7$
 &  ${5.79\cdot 10^{7}}$
 &  $   -0.04\pm0.10$  &  $   25.14_{-25.04}^{+22.10}$  &  $    0.04_{-0.04}^{+0.02}$  &  $    3.44_{-3.44}^{+2.02}$ \\
32& Kepler-145 c  &  $    0.25\pm0.05$  &  $    0.39\pm0.01$  &  $     2.1-     3.1$
 &  ${8.03\cdot 10^{7}}$
 &  $    0.13\pm0.10$  &  $   73.86_{-16.42}^{+17.21}$  &  $    0.92\pm0.03$  &  $   50.39_{-13.43}^{+10.20}$ \\
33$^*$& Kepler-413 b  &  $    0.21\pm0.07$  &  $    0.39\pm0.01$  &  $     0.5-    10.0$
 &  ${3.20\cdot 10^{6}}$
 &  $   -0.20\pm0.20$  &  $   60.44_{-20.85}^{+21.91}$  &  $    0.89_{-0.03}^{+0.04}$  &  $  112.32_{-70.60}^{+40.09}$ \\
34& Kepler-419 b  &  $    2.50\pm0.30$  &  $    0.96\pm0.12$  &  $     1.6-     4.1$
 &  ${4.81\cdot 10^{7}}$
 &  $    0.09\pm0.15$  &  $  198.16_{-100.72}^{+126.80}$  &  $    0.25_{-0.15}^{+0.13}$  &  $   15.41_{-13.45}^{+8.16}$ \\
35& Kepler-420 b  &  $    1.45\pm0.35$  &  $    0.94\pm0.12$  &  $     6.3-    12.3$
 &  ${1.57\cdot 10^{7}}$
 &  $    0.27\pm0.09$  &  $  110.51_{-59.61}^{+88.89}$  &  $    0.24_{-0.16}^{+0.12}$  &  $    9.25_{-7.07}^{+4.90}$ \\
36$^\dagger$& Kepler-432 b  &  $    5.84\pm0.05$  &  $    1.10\pm0.03$  &  $     2.6-     4.2$
 &  ${1.73\cdot 10^{8}}$
 &  $   -0.02\pm0.06$  &  $   67.39_{-67.29}^{+60.95}$  &  $    0.04_{-0.04}^{+0.02}$  &  $    2.74_{-2.74}^{+1.61}$ \\
37& Kepler-539 b  &  $    0.97\pm0.29$  &  $    0.75\pm0.02$  &  $     0.4-     1.2$
 &  ${5.54\cdot 10^{6}}$
 &  $   -0.01\pm0.07$  &  $  154.94_{-48.76}^{+51.90}$  &  $    0.50\pm0.03$  &  $   36.74_{-7.07}^{+5.75}$ \\
38$^*$& Kepler-1647 b  &  $    1.52\pm0.65$  &  $    1.06\pm0.01$  &  $     3.9-     4.9$
 &  ${7.91\cdot 10^{5}}$
 &  $   -0.14\pm0.05$  &  $   29.22_{-15.14}^{+19.76}$  &  $    0.05\pm0.02$  &  $    5.34_{-2.11}^{+2.10}$ \\
39& WASP-8 b  &  $    2.24_{-0.09}^{+0.08}$  &  $    1.04_{-0.05}^{+0.01}$  &  $     3.0-     5.0$
 &  ${1.75\cdot 10^{8}}$
 &  $    0.17\pm0.07$  &  $   84.34_{-38.00}^{+43.33}$  &  $    0.12_{-0.06}^{+0.05}$  &  $    5.79_{-3.27}^{+2.65}$ \\
40& WASP-29 b  &  $    0.24\pm0.02$  &  $    0.79_{-0.04}^{+0.06}$  &  $     7.0-    13.0$
 &  ${2.05\cdot 10^{8}}$
 &  $    0.11\pm0.14$  &  $   25.65_{-6.09}^{+6.63}$  &  $    0.33_{-0.08}^{+0.07}$  &  $   19.28_{-9.52}^{+6.18}$ \\
41& WASP-59 b  &  $    0.86\pm0.04$  &  $    0.78\pm0.07$  &  $     0.1-     1.2$
 &  ${4.41\cdot 10^{7}}$
 &  $   -0.15\pm0.11$  &  $  128.37_{-25.06}^{+26.73}$  &  $    0.47\pm0.09$  &  $   48.78_{-18.26}^{+13.21}$ \\
42$^\dagger$& WASP-69 b  &  $    0.26\pm0.02$  &  $    1.06\pm0.05$  &  $     0.5-     4.0$
 &  ${1.94\cdot 10^{8}}$
 &  $    0.14\pm0.08$  &  $    7.23_{-7.13}^{+4.96}$  &  $    0.09_{-0.09}^{+0.05}$  &  $    4.56_{-4.56}^{+2.56}$ \\
43& WASP-80 b  &  $    0.54\pm0.04$  &  $    1.00\pm0.03$  &  $     0.5-    10.0$
 &  ${1.06\cdot 10^{8}}$
 &  $   -0.13_{-0.17}^{+0.15}$  &  $   20.31_{-7.74}^{+8.47}$  &  $    0.12_{-0.05}^{+0.04}$  &  $   12.28_{-8.82}^{+5.37}$ \\
44& WASP-84 b  &  $    0.69\pm0.03$  &  $    0.94\pm0.02$  &  $     0.4-     1.4$
 &  ${9.26\cdot 10^{7}}$
 &  $    0.00\pm0.10$  &  $   53.82_{-6.48}^{+6.79}$  &  $    0.24\pm0.03$  &  $   17.88_{-5.38}^{+4.00}$ \\
45& WASP-130 b  &  $    1.23\pm0.04$  &  $    0.89\pm0.03$  &  $     2.0-    14.0$
 &  ${1.09\cdot 10^{8}}$
 &  $    0.26\pm0.10$  &  $  108.83_{-17.35}^{+17.58}$  &  $    0.28\pm0.04$  &  $   11.22_{-3.63}^{+2.74}$ \\
46& WASP-132 b  &  $    0.41\pm0.03$  &  $    0.87\pm0.03$  &  $     0.5-    14.0$
 &  ${7.62\cdot 10^{7}}$
 &  $    0.22\pm0.13$  &  $   33.70_{-6.40}^{+6.87}$  &  $    0.26\pm0.05$  &  $   11.63_{-4.92}^{+3.38}$ \\
47& WASP-139 b  &  $    0.12\pm0.02$  &  $    0.80\pm0.05$  &  $     0.3-     1.0$
 &  ${1.61\cdot 10^{8}}$
 &  $    0.20\pm0.09$  &  $   15.64_{-2.48}^{+2.93}$  &  $    0.42\pm0.05$  &  $   19.40_{-5.31}^{+4.07}$ \\
\enddata
\tablecomments{A * indicates circumbinary planets.  A $\dagger$ indicates those planets with results adjusted to reflect the fact that a certain portion of their samples could not be modeled (see \S \ref{Analysis} for discussion). Sources -- 1: \cite{CoRoT-8b}, 2: \cite{CoRoT-9b}, 3: \cite{CoRoT-10b}, 4: \cite{CoRoT-24b}, 5: \cite{Bean2006,Southworth2010}, 6: \cite{HAT-P-11b}, 7: \cite{HAT-P-12b}, 8: \cite{HAT-P-15b}, 9: \cite{HAT-P-17b/c}, 10: \citep{HAT-P-18b/19b}, 11: \cite{HAT-P-20b}, 12: \cite{HAT-P-54b}, 13: \cite{HATS-6b}, 14: \cite{HATS-17b}, 15: \cite{HD17156b}, 16: \cite{HD80606b1} \& \cite{HD80606b2} \& \cite{HD80606b3}, 17: \cite{K2-19}, 18: \cite{K2-24}, 19: \cite{K2-24},  20: \cite{K2-27}, 21: \cite{Kepler-9}, 22: \cite{Kepler-9}, 23: \cite{Kepler-16b}, 24: \cite{Kepler-30}, 25: \cite{Kepler-30}, 26: \cite{Kepler-34b/35b}, 27: \cite{Kepler-34b/35b}, 28: \cite{Kepler-45b}, 29: \cite{Kepler-75b}, 30: \cite{Kepler89d1} \& \cite{Kepler89d2}, 31: \cite{Kepler-117}, 32: \cite{Kepler-145}, 33: \cite{Kepler413b}, 34: \cite{Kepler419b1} \& \cite{Kepler419b2}, 35: \cite{Kepler-420b}, 36: \cite{Kepler-432b}, 37: \cite{Kepler-539}, 38: \cite{Kepler-1647}, 39: \cite{WASP-8b}, 40: \cite{WASP-29b}, 41: \cite{WASP-59b}, 42: \cite{WASP-69b/84b}, 43: \cite{WASP-80b}, 44: \cite{WASP-69b/84b}, 45: \cite{WASP-130/132/139}, 46: \cite{WASP-130/132/139}, 47: \cite{WASP-130/132/139}
}

\end{deluxetable}
\global\pdfpageattr\expandafter{\the\pdfpageattr/Rotate 0}
\clearpage

\bibliographystyle{apj}
\bibliography{references}

\end{document}